\documentclass[journal]{IEEEtran}
\IEEEoverridecommandlockouts

\usepackage{amsmath}
\usepackage{caption}
\usepackage{comment}
\usepackage{amsmath}
\usepackage{mfirstuc}
\usepackage{tabulary,xcolor}
\usepackage{amsfonts,amsmath,amssymb}
\usepackage[T1]{fontenc}
\makeatletter
\newtheorem{theorem}{Theorem}
\usepackage{url,multirow,morefloats,floatflt,cancel,tfrupee}
\makeatother
\usepackage{colortbl}
\usepackage{xcolor}
\usepackage{pifont}
\usepackage[nointegrals]{wasysym}
\makeatletter
\usepackage{float}
\usepackage[utf8]{inputenc}
\usepackage[english]{babel}
\usepackage{blindtext}
\usepackage[utf8]{inputenc}
\usepackage[english]{babel}
\usepackage{fancyhdr}
\usepackage{lastpage}
\usepackage[english]{babel}
\usepackage[utf8]{inputenc}
\usepackage{fancyhdr}
\usepackage{rotating}
\usepackage{color}
\usepackage{diagbox}
\usepackage{graphicx}
\usepackage{adjustbox}
\usepackage{amsmath,amssymb}
\usepackage{mathtools}
\usepackage{amsmath}
\usepackage{subfigure}

\usepackage{cite}
\usepackage{amsmath,amssymb,amsfonts}
\usepackage{algorithmic}
\usepackage{graphicx}
\usepackage{textcomp}
\usepackage{xcolor}
\usepackage{comment}
\usepackage{mfirstuc}
\usepackage{cite}

\def\BibTeX{{\rm B\kern-.05em{\sc i\kern-.025em b}\kern-.08em
    T\kern-.1667em\lower.7ex\hbox{E}\kern-.125emX}}
\usepackage{caption}
\usepackage[utf8]{inputenc}
\usepackage{soul}
\usepackage{amsmath,amssymb,amsfonts}
\usepackage{algorithmic}
\usepackage{textcomp}
\usepackage{xcolor}
\usepackage{comment}
\usepackage{mfirstuc}
\def\BibTeX{{\rm B\kern-.05em{\sc i\kern-.025em b}\kern-.08em
    T\kern-.1667em\lower.7ex\hbox{E}\kern-.125emX}}
\usepackage{mfirstuc}
\usepackage{booktabs}
\usepackage{tabularx}
\usepackage{caption}
\usepackage{xcolor}

\begin{document}


\pagenumbering{arabic}

\thispagestyle{empty}
	
\title{Towards Equitable Peering: A Proposal for a Fair Peering Fee Between ISPs and Content Providers}

\author{Ali~Nikkhah,~\IEEEmembership{Student Member,~IEEE}
and~Scott~Jordan,~\IEEEmembership{Member,~IEEE}\\
\textit{Department of Computer Science, University of California, Irvine, USA} \\
        \{ali.nikkhah, sjordan\}@uci.edu
}

\maketitle
\thispagestyle{empty}

\begin{abstract}

Disagreements over peering fees have risen to the level of potential government regulation. ISPs assert that content providers should pay them based on the volume of downstream traffic. Transit providers and content providers assert that consumers have already paid ISPs to transmit the content they request and that peering agreements should be settlement-free.

Our goal is to determine the fair payment between an ISP and an interconnecting network. We consider fair cost sharing between two Tier-1 ISPs, and derive the peering fee that equalizes their net backbone transportation costs. We then consider fair cost sharing between an ISP and a transit provider. We derive the peering fee that equalizes their net backbone transportation costs, and illustrate how it depends on the traffic ratio and the amount of localization of that content. Finally, we consider the fair peering fee between an ISP and a content provider. We derive the peering fee that results in the same net cost to the ISP, and illustrate how the peering fee depends on the number of interconnection points and the amount of localization of that content. We dispense with the ISP argument that it should be paid regardless of the amount of localization of content.

\end{abstract}


\section{Introduction}

\IEEEPARstart{A}{n} Internet Service Provider (ISP) enables the transmission and receipt of data to and from all or almost all Internet endpoints. To offer this Internet access service, the ISP must establish connections with other networks to exchange data. An interconnection agreement is considered a \textit{transit service} if the transit provider agrees to accept and deliver data on behalf of the ISP, regardless of the destination. On the other hand, if each network agrees to only accept and deliver data with destinations in its customer base, the interconnection agreement is referred to as \textit{peering}. We focus on peering in this paper. Peering may be either paid (i.e., one interconnecting network pays the other) or settlement-free (i.e., without payment).

Recently, it has become a widespread practice for large content providers to directly peer with large ISPs. However, there have been frequent disagreements between them about whether the peering agreement should be paid or settlement-free. The academic literature has not provided much clarity on the appropriate circumstances for settlement-free peering between an ISP and a content provider.

In the United States, between 2013 and 2014, a disagreement between Comcast and Netflix regarding interconnection terms persisted for an extended period. In 2014, Netflix and some transit providers brought the matter to the Federal Communications Commission (FCC) of the United States, which was drafting revised net neutrality regulations at the time. The debate shows contrasting views. Some large content providers and some transit providers claimed that large ISPs created congestion in order to force paid peering arrangements, and that this congestion caused harm to consumers and stifled innovation. They argue that the content providers and transit providers cover the cost of transmitting data to the ISP by localizing content closer to end users, and that the cost to the ISP of adding enough interconnection capacity to ensure minimal congestion is small. In response, large ISPs argue that content providers and transit providers are imposing costs on them as they have to upgrade their infrastructure to handle demand, which raises the bills of all subscribers. They argue that settlement-free peering is a barter arrangement in which each party should receive something of value, and if one party only sends traffic, it is not contributing anything of value.

The FCC addressed the debate surrounding interconnection arrangements in its 2015 Open Internet Order \cite{FCC}, and asserted oversight over interconnection arrangements. Then, in 2018, the FCC reversed its stance, as part of repealing most of the 2015 net neutrality regulations \cite{FCC2}, ending this oversight. However, it is highly likely that the FCC will revisit this issue in the near future as part of a new net neutrality proceeding. In addition, the FCC recently reported that some stakeholders are proposing that content providers pay a fee based on their download traffic to subsidize broadband Internet access in rural areas and for low-income consumers \cite{FCC3}. These advocates are using similar arguments that large ISPs have used to advocate for paid peering.

Similar debates over paid peering are also active in South Korea and in Europe. In South Korea, paid peering between ISPs is now mandatory, based on the amount of traffic exchanged. As a result, these peering fees are often passed on to content providers that interconnect with ISPs in South Korea. A proposal is currently under consideration in South Korea to also require content providers to pay usage fees to ISPs, based on traffic volume \cite{Korea}. The European trade association representing numerous ISPs in Europe has recently put forward a similar proposal, suggesting that content providers should pay usage fees to ISPs, based on the volume of traffic \cite{ETNSO}. However, European regulators are concerned that such fees could be abused by ISPs and are skeptical of the argument that ISPs' costs are not adequately covered by their customers \cite{BEREC}.

In this paper, we address this debate over paid peering fees. Our goal is to determine the fair payment (if any) between an ISP and an interconnecting transit provider or content provider. We define \textit{fair} based on the backbone transportation costs incurred by the ISP and the interconnecting network. When the fair payment is zero, we consider settlement-free peering to be a fair interconnection arrangement. Thus, we are particularly interested in the conditions under which settlement-free peering is fair.

The backbone transportation costs of each interconnecting network is a function of the number of interconnection points, the amount of traffic passing in each direction through interconnection points, and the distance the traffic passes through each network. In turn, these distances are functions of the amount of localization (if any) of content. As video traffic currently constitutes the majority of traffic exchange, and since video is often localized using content delivery networks, backbone transportation costs are a function of video traffic and localization.

The paper is organized as follows. In Section \ref{sec:Literature}, we summarize the relevant research literature. In Section \ref{sec:Model}, we develop a simplified model of backbone transportation costs. We partition an ISP's network into access networks, middle mile networks, and a backbone network. We assume that an ISP and a transit provider or content provider mutually determine a set of points at which to interconnect, chosen from a list of the largest traffic exchange locations in the United States. In order to determine the routes over which traffic flows between networks, we construct traffic matrices, using United States census statistics to determine broadband subscriber locations. We determine the distances on an ISP's backbone network over which it carries traffic to and from a subscriber, and we calculate the average distance using the traffic matrices. We then construct a simplified model of backbone transportation costs as a function of both distance and traffic volume.

In Section \ref{sec:ISP-ISP}, we consider fair cost sharing between two ISPs. We first argue that only backbone transportation costs should be considered in determining fair cost sharing, and that middle-mile and access network transportation costs are appropriately borne by the ISP's subscribers. We briefly examine the traditional settlement-free peering arrangement between a pair of Tier-1 ISPs, and assume that such arrangements reflect fair cost sharing of backbone transportation costs. We then consider the case in which the two interconnecting ISPs have a traffic ratio other than 1. We derive a fair payment between the two ISPs based on the difference in their backbone transportation costs caused by the traffic ratio. 

In Section \ref{sec:TP-ISP}, we turn to the case of an ISP interconnecting with a transit provider. Transit providers increasingly carry not only traffic indirectly passing from one ISP to another, but also content provider traffic. In this case, the ratio of downstream traffic (from the transit provider to the ISP) to upstream traffic (from the ISP to the transit provider) is likely to be higher than when two ISPs interconnect, because the content provider traffic is almost entirely downstream video traffic. The higher traffic ratio increases the ISP's backbone transportation costs. However, the transit provider may deliver a portion of the video traffic using cold potato routing, which localizes traffic on the ISP's network and reduces the ISP's backbone transportation costs. We model both the ISP's and the transit provider's backbone transportation costs, as a function of the number of interconnection points and the traffic ratio between the two, routing, and localization.

In Section \ref{sec:Indirect}, we consider fair cost sharing between an ISP and a transit provider. We first consider the case in which the transit provider uses hot potato routing for all traffic. We derive the peering fee that equalizes the ISP's and the transit provider's net costs, and show that it is similarly a function of the difference in their backbone transportation costs caused by an unequal traffic ratio. We then consider the case in which the transit provider uses cold potato routing for a proportion of the video traffic. We again derive the peering fee that equalizes the ISP's and the transit provider's net costs, and show how it depends not only on the traffic ratio, but also on the amount of video traffic and the amount of video content localization. The transit provider should pay the ISP for peering if it doesn't localize a sufficient portion of the video traffic. The fair peering fee may be positive and substantial if there is a high volume of video traffic with low localization. Finally, we consider the case in which the transit provider uses a CDN to localize traffic instead of delivering it using cold potato routing. We argue that the fair peering fee should be unchanged, and that a CDN will result in cost savings if the cost of building it is less than the cost of carrying traffic across the transit provider's backbone.

In Section \ref{sec:Direct}, we consider fair cost sharing between an ISP and a directly interconnected content provider. We argue that an ISP should be indifferent between peering with another ISP or a transit provider and peering directly with a content provider, if the sum of the ISP's backbone transportation costs and any peering fee is unaffected. We discuss the implications of this equivalence. We show that the fair peering fee depends on the localization of video traffic and the number of interconnection points. If the content provider does not localize enough video traffic, it should pay the ISP a peering fee. As the content provider localizes more video traffic, the peering fee should decrease.

Finally, we re-examine the arguments put forth by large ISPs and large content providers. We reject ISP assertions that they should apply the same settlement-free peering requirements to both peering ISPs and peering content providers. We also reject ISP assertions that they should be compensated by large content providers regardless of the amount of video content localization. We also reject any assertions by transit providers or content providers that should be entitled to settlement-free peering solely because the ISP's customers have already paid the ISP to transport the traffic the content providers are sending. Instead, we argue that the settlement-free peering requirements for content providers should include a specified minimum number of interconnection points and a specified minimum amount of traffic to be delivered locally.

\section{Research Literature}\label{sec:Literature}

The academic literature has not provided much clarity on the appropriate circumstances for settlement-free peering between an ISP and a content provider. In \cite{us_journal}, we constructed a network cost model to understand the rationality of common peering requirements, including the number of interconnection points, routing, and localization. However, we did not determine what constitutes a fair peering fee.

There are a few papers that address the issue of fair cost sharing between ISPs. Gyarmati et al. \cite{gyarmati2012sharing} consider multiple ISPs transmitting traffic over a transit provider's network. They examine various mappings from usage to cost. They show some mappings can achieve a fair and efficient allocation of costs among ISPs, while also providing incentives for network investment and capacity planning. Although they don't discuss peering fees, their results could be used to assign fair peering fees between an ISP and multiple content providers based on the traffic of content providers that passes over the ISP's backbone network. However, since they do not explicitly model peering between various parties, their estimates of cost do not explicitly consider the number of interconnection points or the localization of traffic. 

There are some papers that focus on the economics of Internet interconnection, which involves the mechanisms and incentives for ISPs and other networks to connect with each other for the exchange of Internet traffic. Dovrolis \cite{dovrolis2015evolution} and Tan et al. \cite{tan2006economic} delve into the historical development of Internet interconnection, including the rise of content delivery networks (CDNs) and the emergence of settlement-free peering. Ma \cite{ma2017pay} and Wang et al. \cite{wang2021paid} examine the effects of different peering arrangements on Internet traffic, including the impact on network performance and congestion. Patchala et al. \cite{patchala2021economics} study the effects of net neutrality regulations on interconnection arrangements, including the impact on traffic routing and the financial implications for ISPs. Courcoubetis et al. \cite{courcoubetis2016negotiating} and Zarchy et al. \cite{zarchy2018nash} present various models for calculating peering prices and for evaluating the benefits of different peering strategies. Dhamdhere et al. \cite{dhamdhere2010value} and Ma \cite{ma2020internet} advocate for using a value-based framework that takes into account the mutual benefits of peering arrangements, rather than relying solely on market-based pricing mechanisms. These papers provide insights into the economic factors that influence Internet interconnection and the challenges in managing the interconnection ecosystem. However, they do not take into account the impact of traffic ratios and content localization on ISPs' backbone costs, nor do they calculate a fair peering fee between ISPs and transit or content providers. 

Another group of papers focus on comparing peering and transit interconnections in terms of performance and cost reduction. Castro and Gorinsky \cite{castro2012t4p} propose a hybrid peering model for interconnecting transit providers and ISPs that reduces backbone transport costs. However, they focus on cost reduction for transit providers and ISPs, while we focus here on determining a fair payment for peering arrangements. Ahmed et al. \cite{ahmed2017peering} compare the performance of peering and transit interconnection. However, they do not consider the economic aspects of interconnection such as cost sharing, we focus on determining the fair payment between an ISP and an interconnecting transit provider or content provider for peering. Dey and Yuksel \cite{dey2019peering} compare the performance of different peering scenarios, including direct peering, public peering, and paid peering. However, they focus on the peering strategies of vertically integrated ISPs that provide both content and access services, while we focus on determining the fair payment between ISPs and content providers or transit providers for peering arrangements. Overall, the papers in this group contribute to our understanding of the trade-offs between different interconnection models and the factors that influence their performance and cost effectiveness.

Another group of papers focus on CDNs and their role in Internet interconnection. Böttger et al. \cite{bottger2018open} provide an analysis of the Netflix CDN and its impact on the Internet ecosystem, including the potential for optimizing server placement, increasing the use of settlement-free peering, and exploring new interconnection models. Netflix itself describes its approach to working with network operators and content providers to improve the performance and efficiency of content delivery in \cite{Netflix}. These papers shed light on the role of CDNs in Internet interconnection and the implications for the Internet ecosystem.

Another group of papers focuses on paid peering, which refers to the situation where a content provider pays an ISP for the delivery of its traffic. Jitsuzumi \cite{jitsuzumi2022economic} discusses a lawsuit in South Korea by Netflix against SK Broadband regarding peering fees. He presents an analysis that shows that paid peering is neutral to resource allocation when pricing is not constrained, but beneficial to ISPs and their subscribers when pricing is constrained. Lee et al. \cite{lee2022economics} analyze the economic impact of CDNs on content providers and show that CDNs can help content providers reduce their delivery costs and improve their quality of service but also raise concerns about the potential for market concentration and monopoly power. Wang and Ma \cite{wang2020private} analyze the optimal pricing and contract terms for direct peering agreements between content providers and ISPs, and show that direct peering can be mutually beneficial for both parties but the pricing and contract terms should be carefully negotiated to ensure fairness and efficiency. These papers provide insights into the economic and technical factors that influence paid peering and the challenges in managing the interconnection relationships between content providers and ISPs.

\section{Model}\label{sec:Model}

In this section, we develop a simplified model of backbone transportation costs in the United States.\footnote{Outside the United States, other models may be more appropriate given differences in network topology.} The goal of the model is to analyze cost sharing, and in particular the dependence of network costs on traffic. We focus on the characteristics that are most critical to this analysis, and abstract less critical characteristics. We recognize that precise network costs will differ from those derived using this simplified model, but we believe that the simplified model is sufficient to illustrate the dependence of fair peering fees on the number of interconnection points, the traffic ratio, the amount of downstream content, and the amount of localization of that content.

We consider an ISP that serves customers throughout the contiguous United States. The ISP's network is partitioned into a backbone network, middle mile networks, and access networks. Section \ref{sec:topology} presents this topology. 

We then model traffic matrices by assuming traffic is proportional to the population. We consider traffic exchange between an ISP and a transit provider or content provider at the geographic locations of the largest interconnection points (IXPs) in the United States. Section \ref{sec:traffic} develops the traffic matrices.

We then model backbone transportation costs. We model the traffic-sensitive cost associated with carrying the traffic over the backbone, as a function of routing and distances. Section \ref{Sec:traffic_costs} presents models of these distances, given traffic matrices. To help readers easily refer to the symbols used in this paper, we provide a glossary of symbols in Table \ref{tab:symbols}. This table includes all the symbols used in the paper and their corresponding descriptions.

\begin{table}
  \centering
  \caption{Glossary of Symbols}
  \label{tab:symbols}
  \begin{tabular}{@{}ll@{}}
    \toprule
    Symbol & Description \\
    \midrule
    $A(j)$ & Geographical center of access network $j$ \\
    $Access(j)$ & Geographical region of access network $j$ \\
    $C_{ISP}$ & ISP backbone cost\\
    $C_{TP}$ & Transit provider backbone cost\\
    $C^{ISP}_{cp,video}$ & ISP backbone cost when peering with content provider\\
    $c^b$ & Cost per unit distance and volume in backbone network\\
    $D^b$ & Distance on backbone network\\
    $ED$ & Average distance on backbone network\\
    $IXP(i)$ & Location of IXP $i$ \\
    $IXP^{u}$ & Location of the IXP closest to the end user \\
    $IXP^{p}$ & Location of the IXP at which traffic enters/exits the ISP \\
    $I$ & Set of locations of the IXPs \\
    $l^N$ & Set of $N$ IXPs at which the ISP agrees to peer \\
    $L$ & Distance from west to east of the United States \\
    $M$ & Number of major IXPs \\
    $N$ & Number of IXPs at which the ISP agrees to peer \\
    $p$ & Population of the contiguous United States \\
    $P(j)$ & Probability that an end user resides in access network $j$ \\
    $p_j$ & Population of the county associated with access network $j$ \\
    $P^{TP,ISP}$ & Fair peering fee between transit provider and ISP\\
    $P_{v}^{TP,ISP}$ & Component of $P^{TP,ISP}$ related only to video traffic\\
    $P_{CP,ISP}$ & Fair peering fee between content provider and ISP\\
    $r$ & Ratio of non-video downstream to upstream traffic \\
    $r'$ & Ratio of video downstream to upstream traffic\\
    $R(IXP(i))$ & Geographical region of IXP $i$'s access networks \\
    $s_j$ & Size of county $j$ \\
    $S$ & Traffic source's location \\
    $U$ & End user's location \\
    $V_d$ & Volume of non-video downstream traffic\\
    $V_v$ & Volume of video downstream traffic\\
    $V_u$ & Volume of upstream traffic\\
    $x$ & Fraction of video traffic localized by transit provider\\
    $x_d$ & Fraction of video traffic localized by content provider\\
    \bottomrule
  \end{tabular}
\end{table}

\subsection{Topology} \label{sec:topology}
Our goal is to analyze the traffic-sensitive backbone transportation costs incurred by the ISP when interconnecting with another network. These costs depend on the average distance that traffic travels over the ISP's backbone, which in turn depends on routing and traffic demand patterns. To calculate the average backbone distances, we will model the ISP's service territory and construct traffic matrices representing subscriber locations. We will use these distances in Section III-C to derive a simplified model of the ISP's traffic-sensitive backbone transportation costs as a function of both distance and traffic volume.

We model the ISP's network as partitioned into a single backbone network, multiple middle-mile networks, and multiple access networks. The backbone network is assumed to connect all of the IXPs at which the ISP is present. A middle-mile link is assumed to run from the geographical center of each access network to the closest IXP, and each access network is assumed to span a single U.S. county. While we recognize that topologies of access networks differ widely, this assumption will not affect the results in this paper, since peering policies depend more critically on the location and number of interconnection points than on the topologies of access networks.

Our model of the topology of an ISP's network consists of the ISP's service territory, the location of Internet Exchange Points (IXPs), and the segments of its network.

We consider an ISP whose service territory covers the contiguous United States.\footnote{While most ISPs do not offer residential broadband Internet access service over the entire contiguous United States, we see little in their settlement-free peering policies that are specific to their service territory, other than that a subset of the interconnection points at which they peer are concentrated near their service territory.} We represent this service territory using the set of longitudes and latitudes of the contiguous United States.

We consider $M$ IXPs at which an ISP may agree to peer with another ISP, a transit provider, or a content provider. In the numerical results below, we use the $M=12$ largest IXPs in the United States, located at Ashburn, Chicago, Dallas, San Jose, Los Angeles, New York, Seattle, Miami, Atlanta, Denver, Boston, and Minneapolis \cite{PeeringDB}. Denote the coordinates (in longitude and latitude) of these $M$ IXPs by $IXP(i)$, and the set of the locations of these IXPs by $I^M$. An ISP and a content provider often agree to interconnect at a smaller number $N<M$ of IXPs. Denote the set of $N$ IXPs at which they agree to interconnect by $l^N \subseteq \{1,...,M\}$, and the set of locations of these IXPs by $I^N = \{IXP(i),i \in l^N\} \subseteq I^M$.

Denote the geographical region of access network $j$ by $Access(j)$, and the location of the geographical center of access network $j$ by $A(j)$. We assign these locations using the longitudes and latitudes of the center of each county in the contiguous United States.

Consider an ISP and an interconnecting network that agree to interconnect at the $N$ IXPs in $l^N$. Denote by $R^N(IXP(i))$ the geographical region that consists of the union of access networks for which the closest IXP in $l^N$ is IXP $i$, namely 
\begin{equation}
R^N(IXP(i)) = \bigcup\limits_{\substack{j \, \mid \, \lVert A(j)-IXP(i) \rVert \, \leq \, \\ \lVert A(j)-IXP(i') \rVert \, \forall i' \in l^N}} Access(j)
\end{equation}

\subsection{Traffic matrices} \label{sec:traffic}

\begin{figure}
\centering
\includegraphics[width=\columnwidth]{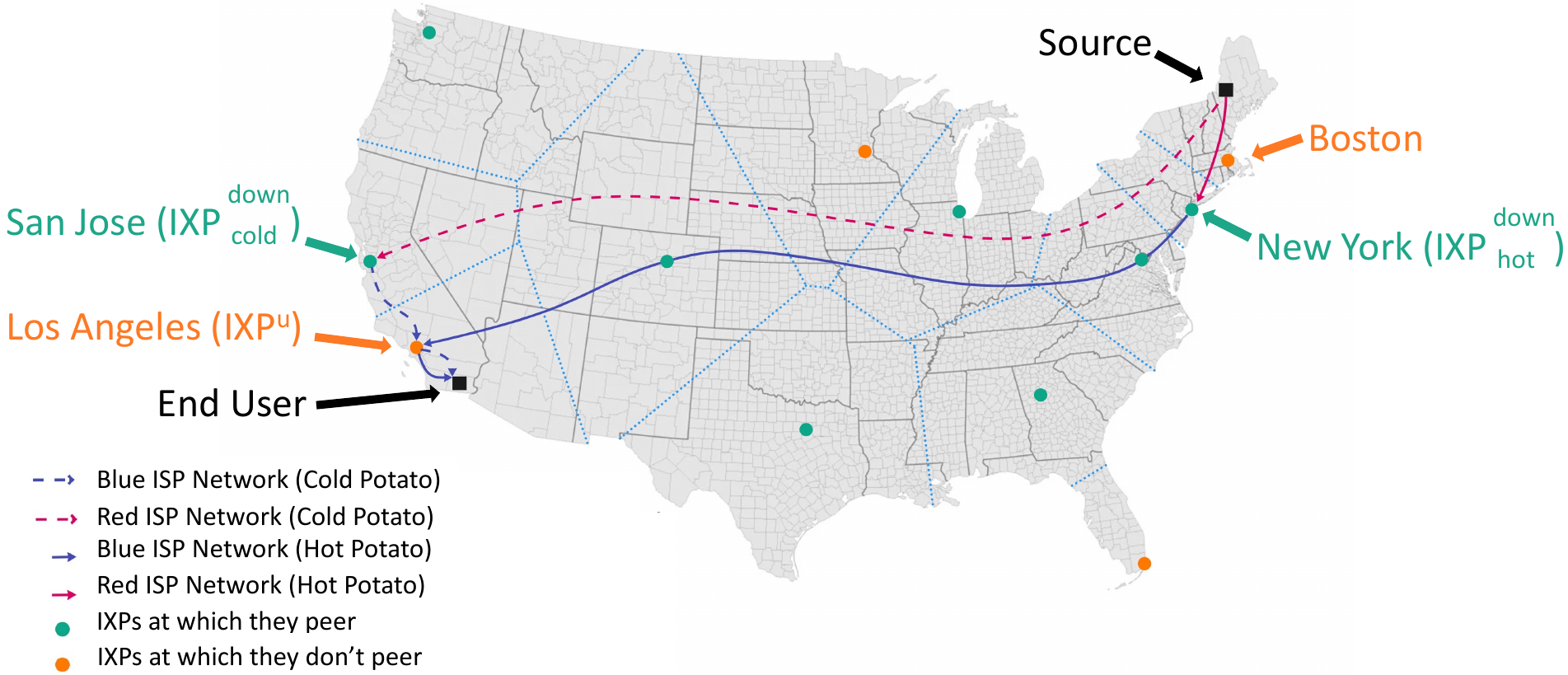}
\caption{Topology of an ISP's network}
\label{fig:routing}
\end{figure}

The locations of end users of the ISP are represented by a probability distribution over the ISP's service territory. We decompose this distribution into (a) a distribution of the number of end users in each access network and (b) for each access network, the distribution of end users within the access network.

Denote the probability that an end user resides within access network $j$ by $P(j)$. We assume that end users are distributed across access networks according to the population of the county associated with the access network, denoted by $p_j$, and we denote the population of the contiguous United States by $p = \sum_j p(j)$. These populations are taken from U.S. census data \cite{Census}. It follows that $P(j) = p_j/p$. We further assume that end users are uniformly distributed within each access network, and we denote the size of county $j$ by $s_j$, which we determine using the U.S. Gazetteer \cite{Lat_Long}.

We focus first on downstream traffic that originates outside the ISP's network and terminates at an end user on the ISP's network. Denote the source's location by $S$ and the end user's location by $U$. We assume that $S$ and $U$ are independent and that the marginal distributions of $S$ and $U$ are given by the joint distribution of the population with each access network and the distribution of end users within each access network.

In this paper, the terms "hot potato" routing and "cold potato" routing refer to the routing decisions made by the network provider where the originating source of the traffic is located. 

Along the route $S$ to $U$, denote the location of the IXP at which downstream traffic using hot potato routing enters the ISP's network by $IXP_{hot}^{down}$, the location of the IXP at which downstream traffic using cold potato routing enters the ISP's network by $IXP_{cold}^{down}$, and the location of the IXP closest to the end user by $IXP^u$. For example, Figure \ref{fig:routing} provides a rough illustration of a scenario where the ISP hosting the end user (blue ISP) and the interconnecting network (red ISP), where the source is located, agree to interconnect at $N=8$ IXPs. Suppose $S$ is in Maine and $U$ is in Imperial county, California. Then, as illustrated in Figure \ref{fig:routing}, $IXP_{hot}^{down}$ might be in New York (if the two networks do not agree to peer in Boston), $IXP_{cold}^{down}$ might be in San Jose (if the two networks do not agree to peer in Los Angeles), and $IXP^u$ is in Los Angeles.

The ISP carries traffic on only part of the route from $S$ to $U$. It carries traffic on its backbone from $IXP_{hot}^{down}$ to $IXP^u$, and it carries traffic on a middle mile network and an access network from $IXP^u$ to $U$. The part of the route that is on the ISP's network thus depends on the joint distribution of $(IXP_{hot}^{down},IXP^u,U)$.

The access network on which $U$ resides is distributed according to $\{P(j)\}$, and $U$ is uniformly distributed within the access network. The IXP closest to the end user is a deterministic function of $U$, namely $IXP^u=(g' | \, U \in R^M(g'))$. 

The IXP at which downstream traffic using hot potato routing enters the ISP's network ($IXP_{hot}^{down}$) is independent of the end user, and it is the IXP closest to the source among the IXPs at which they agree to peer, i.e., $IXP_{hot}^{down}=(g | \, S \in R^N(g))$. Since end users are assumed to be distributed according to U.S. county population statistics:
\begin{equation}\label{pmf_s2}
\begin{aligned}
P(IXP_{hot}^{down} = g) = \frac{1}{p}\sum_{Access(j) \subset R^N(g)}p(j)
\end{aligned}
\end{equation}

In contrast, if the ISP and the interconnecting network use cold potato routing, then the IXP at which downstream traffic enters the ISP's network ($IXP_{cold}^{down}$) is no longer independent of the end user, and it is the IXP closest to the end user at which they agree to peer, i.e., $IXP_{cold}^{down}=(g | \, U \in R^N(g))$. 

For upstream traffic, the routes and distributions are similar but inverted. If the ISP and the interconnecting network use hot potato routing, then the IXP at which upstream traffic enters the interconnecting network is the IXP closest to the end user at which they agree to peer, i.e., $IXP_{hot}^{up}=IXP_{cold}^{down}=(g | \, U \in R^N(g))$. If the ISP and the interconnecting network use cold potato routing, then the IXP at which upstream traffic enters the interconnecting network is independent of the end user and follows a distribution similar to (\ref{pmf_s2}).

\subsection{Traffic-sensitive backbone costs} \label{Sec:traffic_costs}

We now construct a simplified model of backbone transportation costs. These costs are a function of the average distance carried over a network's backbone, which in turn depends on routing and the traffic matrices. These distances were derived in \cite{us_ccnc,globe_us}, but we summarize them for convenience. 

For downstream traffic, the distance from $S$ to $U$ on the ISP's backbone network is a function of the location of the IXP at which downstream traffic enters the ISP's network ($IXP_{hot}^{down}$) and the location of the IXP closest to the end user ($IXP^u$). Denote the distance on the ISP's backbone network between these two IXPs by $D^b(IXP_{hot}^{down},IXP^u) = \lVert IXP_{hot}^{down}-IXP^u \rVert$. We separately consider downstream traffic using hot potato routing and downstream traffic using cold potato routing.

For downstream traffic using hot potato routing, the IXP at which the traffic enters the ISP's network ($IXP_{hot}^{down}$) depends on the IXPs at which they agree to interconnect. However, it is independent of the end user and thus independent of the IXP closest to the end user ($IXP^u$). Consider an ISP and an interconnecting network that agree to interconnect at the $N$ IXPs in $I^N$. The average distance of the downstream traffic on the ISP's backbone network, which flows from the source $S$ to the user $U$, is:
\begin{equation}\label{EB2_hot}
\begin{aligned}
ED^{hot}_{down} (N)
&= \sum\limits_{g \in I^N} \sum\limits_{g' \in I^M}\\
&{D^b(g,g') P(IXP_{hot}^{down}=g) P(IXP^u=g')}
\end{aligned}
\end{equation}

The probability distribution of $IXP_{hot}^{down}$ was given in (\ref{pmf_s2}). The probability distribution of $IXP^u$ can be similarly represented as:
\begin{equation}
\begin{aligned}
P(IXP^u=g')=\frac{1}{p}\sum_{Access(j) \subset R^M(g')}p(j)
\end{aligned}
\end{equation}

For downstream traffic using cold potato routing, the IXP at which the traffic enters the ISP's network is the IXP closest to the end user at which they agree to peer, i.e., $IXP_{cold}^{down}$. The ISP might still carry traffic across a portion of its backbone, namely from $IXP_{cold}^{down}$ to $IXP^u$, and the average such distance, which flows from the source $S$ to the user $U$, is:
\begin{equation}\label{EB2_cold_rev}
\begin{aligned}
    ED^{cold}_{down} (N)
    &= \sum_{g' \in I^M} \sum_{g \in I^N}\\
    &{D^b(g,g') P(IXP_{\text{cold}}^{\text{down}}=g, IXP^u=g')}
  \end{aligned}
\end{equation}

Now, since for each $g'$, there exists a unique $g$ which minimizes the backbone distance $D^b(g, g')$, and this $g$ is chosen based on the condition $ g' \in R^N(g)$, we can simplify Equation (\ref{EB2_cold_rev}). In this context, $R^N(g)$ denotes the set of IXPs at which the ISP agrees to peer, given that the traffic has entered the network at IXP $g$.

With this assumption, for each $g'$, we select the $g$ that minimizes $D^b(g, g')$. Consequently, the inner sum in Equation (\ref{EB2_cold_rev}) collapses to a single term for each $g'$, and the equation simplifies to:

\begin{equation}\label{EB2_cold}
ED^{cold}_{down} (N)= \sum\limits_{g' \in I^M} {D^b(g \mid g' \in R^N(g),g') P(IXP^u=g')}
\end{equation}

We are concerned only with the portion of an ISP's backbone transportation costs that is sensitive to the amount of traffic, because non-traffic-sensitive costs do not vary significantly with the number of IXPs or the traffic ratio. Traffic-sensitive costs are a function of both distance and traffic volume. We model traffic-sensitive costs as linearly proportional to the average distance over which the traffic is carried on each portion of the ISP's network \cite{valancius2011many}, and linearly proportional to the average volume of traffic that an ISP carries on each portion of its network. 

Denote the cost per unit distance and per unit volume in the backbone network by $c^b$. Denote the volume of traffic by $V$. The ISP's traffic-sensitive backbone cost is thus $V {c^b} ED^{hot}_{down}$ for downstream traffic using hot potato routing and $V {c^b} ED^{cold}_{down}$ for downstream traffic using cold potato routing. The ISP's traffic-sensitive backbone cost for upstream traffic can be similarly calculated.

\section{Peering between two ISPs}\label{sec:ISP-ISP}

In this section, we analyze peering between two ISPs. We propose that customers bear middle-mile and access costs, and that peering fees result in fair cost sharing between the two ISPs of backbone costs. We calculate the peering fee necessary for equitable cost-sharing and investigate how traffic ratios affect the costs and payments between the two parties.

We first consider peering between two ISPs of comparable size and traffic, i.e., the same number of customers, similar backbone sizes, and the same amount of upload and download traffic. The results from our studies in \cite{us_journal,tprc_us} indicate that as long as there is symmetry in the arrangement, these ISPs would likely reach a settlement-free peering agreement.

Next, we consider peering between two ISPs that carry unequal traffic. In this case, there may be payment between the two networks. The payment will depend on the traffic ratio, as we will analyze in this section.

In this section, we assume that the ISPs share a similar network topology. This means that the mathematical models and equations presented are applicable from the perspective of any of the involved ISPs.

\subsection{Cost Recovery}

When analyzing fair peering fees, it is necessary to define what we mean by \textit{fair}. A large portion of an ISP's costs are recovered from its subscribers. An ISP may, however, also recover some of its costs from interconnecting networks. 

Our focus here is on traffic-sensitive costs, since the debate over paid peering centers on costs incurred because of traffic. However, we must still address whether an ISP should recover traffic-sensitive costs across different parts of its network solely from its subscribers or also from interconnecting networks.

Economists often debate about the proper amount of cost recovery from each side in two-sided markets. In the context of peering, however, there is general agreement that subscribers cover, at a minimum, the costs of an ISP's access and middle-mile networks. The debate is generally over what portion of the costs of an ISP's backbone networks should be borne by subscribers versus interconnecting networks. 

There are several rationales for this approach. First, regulatory cost accounting often dictates that access network costs be recovered from subscribers. Second, the conditions under which two ISPs peer (including routing, number of interconnection points, and traffic ratios) affect ISPs' backbone transportation costs. However, these same conditions do not affect ISPs' middle-mile or access network transportation costs, since an ISP must carry traffic across these portions of its network regardless.

In the remainder of the paper, we thus focus on traffic-sensitive backbone costs.

\subsection{Traffic-sensitive Backbone Cost}

In order to understand the effect of routing policies, traffic ratios, and traffic localization on peering agreements between two ISPs, we first analyze traffic-sensitive backbone costs.

When we refer to "traffic localization," we are describing the strategy wherein the peering network directs traffic to the IXP closest to the end user. This approach ensures that the ISP often avoids transporting the traffic across its backbone network, especially when they have a peering agreement at the nearest IXP to the end user. Even in situations where they don't peer at the closest IXP, the distance the ISP needs to cover to transport the traffic over its backbone is significantly reduced.

We consider two ISPs, denoted as $ISP^1$ hosting the end user ($U$) and $ISP^2$ where the source ($S$) is located. We assume that the two ISPs interconnect at the 12 major interconnection points, and that both use hot potato routing. We consider downstream traffic originating on $ISP^2$'s network destined for an end user located in $ISP^1$'s network\footnote{We consider endpoints in an ISP's customer cone as equivalent to endpoints on an ISP's network, and do not explicitly consider payments between an ISP and it's transit customers.}, and denote the volume of this traffic by $V_d^1$.  We also consider upstream traffic originating with an end user located in $ISP^1$'s network and destined for a location on $ISP^2$'s network, and denote the volume of this traffic by $V_u^1$. We denote the traffic ratio by $r^1=\frac{V_d^1}{V_u^1}$.

We denote $ISP^1$'s traffic-sensitive backbone cost by $C^{ISP^1}$, and partition it into the cost of delivering downstream traffic, which flows from the source $S$ to the user $U$, denoted by  $C^{ISP^1}_{S,U}$; and the cost of delivering upstream traffic, which flows from the user $U$ to the source $S$, denoted by $C^{ISP^1}_{U,S}$:
\begin{equation}\label{C_isp_eq*}
\begin{aligned}
C^{ISP^1}
&= C^{ISP^1}_{S,U}+ C^{ISP^1}_{U,S}
\end{aligned}
\end{equation}

The cost of delivering downstream traffic using hot potato routing is: 
\begin{equation}\label{isp_video*}
\begin{aligned}
C^{ISP^1}_{S,U}= c^b V_d^1 ED^{hot}_{down}(M),
\end{aligned}
\end{equation}
where $c^b$ is the cost per unit distance and per unit volume in the backbone network, and $ED^{hot}_{down}(M)$ is the average distance on $ISP^1$'s backbone network of downstream traffic with hot potato routing, when interconnecting at $M=12$ IXPs.

The cost of delivering upstream traffic using hot potato routing is: 
\begin{equation}\label{isp_video_up}
\begin{aligned}
C^{ISP^1}_{U,S}
&= c^b V_u^1 ED^{hot}_{up}(M)\\
&=c^b V_u^1 ED^{cold}_{down}(M),
\end{aligned}
\end{equation}
because the average distance on $ISP^1$'s backbone network of upstream traffic with hot potato routing, when interconnecting at $M=12$ IXPs, is the same as the average distance on $ISP^1$'s backbone network of downstream traffic with cold potato routing (i.e., $ED^{hot}_{up}(M)=ED^{cold}_{down}(M)$).

Since the two ISPs are assumed to interconnect at all $M=12$ major interconnection points and are assumed to use hot potato routing for upstream traffic, it follows that $ED^{hot}_{up}(M)=ED^{cold}_{down}(M)=0$. This occurs because, within our model's assumptions, when ISPs interconnect at all M IXPs and apply hot potato routing for upstream traffic, they effectively do not carry the traffic across their own backbone network. The traffic is instead offloaded immediately at the closest peering point, which results in no additional distance being traversed on the ISP's backbone network.

As a result, equations (\ref{C_isp_eq*})-(\ref{isp_video_up}) can be simplified to:
\begin{equation}\label{c_isp_isp1}
\begin{aligned}
C^{ISP^1} 
&=c^b V_d^1 ED^{hot}_{down}(M)
\end{aligned}
\end{equation}

Using similar calculations and the definition of the traffic ratio $r^1$, the backbone cost of $ISP^2$ is:
\begin{equation}\label{c_isp_isp2}
\begin{aligned}
C^{ISP^2} 
&=c^b V_d^2 ED^{hot}_{down}(M) \\
&= c^b V_u^1 ED^{hot}_{down}(M)=c^b \frac{V_d^1}{r^1} ED^{hot}_{down}(M)
\end{aligned}
\end{equation}

\subsection{Fair Peering Fee}

We first examine the conditions under which the two ISPs would agree to settlement-free peering. We assume here that they will agree to settlement-free peering if and only if they incur the same amount of traffic-sensitive backbone costs.\footnote{That said, we note that two ISPs may agree to settlement-free peering only if they obtain roughly equal value from the arrangement, and value may not be dictated solely by cost sharing. In particular, we do not consider market power.} Not surprisingly, the two ISPs incur the same backbone cost (i.e., $C^{ISP^1}=C^{ISP^2}$) if and only if the traffic ratio is 1.

We next examine the fair peering fee when the traffic ratio is not 1. Now, in order to equalize net costs, the ISP with a lower traffic-sensitive backbone cost should compensate the other ISP. Denote the fee that $ISP^2$ pays $ISP^1$ for peering by $P^{ISP^2,ISP^1}$. The peering fee that equalizes net costs is given by:
\begin{equation}
\begin{aligned}
C^{ISP^1}-P^{ISP^2,ISP^1}=C^{ISP^2}+P^{ISP^2,ISP^1},
\end{aligned}
\end{equation}
i.e.,
\begin{equation}\label{eq:isp_pay}
\begin{aligned}
P^{ISP^2,ISP^1}=\frac{1}{2} (C^{ISP^1}-C^{ISP^2})
\end{aligned}
\end{equation}

By using equations (\ref{c_isp_isp1}),(\ref{c_isp_isp2}), and (\ref{eq:isp_pay}) we can express the fair peering fee in terms of traffic volumes and average backbone distances:
\begin{equation}
\begin{aligned}
P^{ISP^2,ISP^1}=\frac{1}{2} c^b (V^1_d-V^1_u) ED^{hot}_{down}(M),
\end{aligned}
\end{equation}
or, equivalently, in terms of the traffic ratio $r^1$:
\begin{equation}\label{eq:pay_isp}
\begin{aligned}
P^{ISP^2,ISP^1}=\frac{1}{2} c^b V^1_u (r^1-1) ED^{hot}_{down}(M)
\end{aligned}
\end{equation}

\begin{theorem}\label{theorem:p_isp_isp}
The fair peering fee between two ISPs is: 
\begin{equation}\label{p_isp_isp}
\begin{aligned}
P^{ISP^2,ISP^1}=\frac{1}{2} c^b (V^1_d-V^1_u) ED^{hot}_{down}(M)
\end{aligned}
\end{equation}
\end{theorem}

Theorem \ref{theorem:p_isp_isp} states that the peering fee that equalizes net costs is one half of the difference between the costs incurred by the two ISPs. If the traffic ratio is greater than 1, then the fair peering fee is positive, and if the traffic ratio is less than 1, then the fair peering fee is negative. However, when the traffic ratio is close to 1 (e.g., between 0.5 and 2), then the fair peering fee may be small, and hence the ISPs may choose to adopt settlement-free peering regardless.

\section{Backbone Costs for Peering Between a Transit Provider and an ISP} \label{sec:TP-ISP}

We now turn to peering between a transit provider and an ISP, i.e., neither is a customer of the other. In this section, we consider the traffic-sensitive backbone costs of each. In the following section, we determine the fair peering fee. 

Consider the case in which neither the ISP nor the transit provider is a customer of the other. Rather, they agree to peer with each other. Peering between a transit provider and an ISP is different than peering between two ISPs for two reasons. First, transit providers increasingly carry not only traffic indirectly passing from one ISP to another, but also content provider traffic. In this case, the ratio of downstream traffic (from the transit provider to the ISP) to upstream traffic (from the ISP to the transit provider) is likely to be higher than when two ISPs interconnect, because the content provider traffic is almost entirely downstream video traffic. The higher traffic ratio increases the ISP's backbone transportation costs. 

Second, the transit provider may deliver a portion of the video traffic using cold potato routing, which localizes traffic on the ISP's network and reduces the ISP's backbone transportation costs. Cold potato routing is one such strategy that is commonly employed by transit providers, particularly when delivering video content. This routing method allows the transit provider to retain control over a larger portion of the traffic's journey, potentially leading to improvements in Quality of Service (QoS).

In this section, we model both the ISP's and the transit provider's backbone transportation costs, as a function of the number of interconnection points and the traffic ratio between the two, routing, and localization.  We wish to understand how traffic ratios and video traffic localization could impact the backbone cost of each network.

\subsection{Localization and Routing}

We partition traffic exchanged between a transit provider and an ISP into video traffic and non-video traffic. As before, we assume that non-video traffic is transported using hot potato routing. 

However, for video traffic, we assume that a portion is delivered from the transit provider to the ISP using cold potato routing. Specifically, we assume that a proportion $x$ of the video traffic transmitted to the ISP's users within each access network is delivered from the transit provider using cold potato routing. We assume that the transit provider and the ISP interconnect at all $M=12$ major interconnection points, and consequently, that video traffic delivered using cold potato routing is handed off from the transit provider to the ISP at the IXP closest to the end user.  We assume that the remaining proportion $1-x$ of the video traffic transmitted to the ISP's users within each access network is delivered using hot potato routing, and that the source of this video traffic is independent of the location of the end user.

\subsection{ISP Cost}

We consider downstream traffic destined for an end user located in the ISP's network.  We denote the volume of non-video downstream traffic by $V_d$ and the volume of video downstream traffic by $V_v$.  We also consider upstream traffic originating with an end user located in the ISP's network, and denote the volume of this traffic by $V_u$. 

We define two traffic ratios: $r=\frac{V_d}{V_u}$, the ratio of downstream non-video traffic to upstream traffic, and $r'=\frac{V_v}{V_u}$, the ratio of downstream video traffic to upstream traffic.  

We denote the ISP's traffic-sensitive backbone cost by $C^{ISP}$, and partition it into the cost of delivering downstream non-video traffic, which flows from the source $S$ to the user $U$, (denoted by $C^{ISP}_{S,U,non-video}$), the cost of delivering downstream video traffic, which flows from the source $S$ to the user $U$, (denoted by $C^{ISP}_{S,U,video}$), and the cost of delivering upstream traffic, which flows from the user $U$ to the source $S$, (denoted by $C^{ISP}_{U,S}$):
\begin{equation}\label{C_isp_eq}
\begin{aligned}
C^{ISP} = C^{ISP}_{S,U,non-video}+C^{ISP}_{S,U,video} + C^{ISP}_{U,S}
\end{aligned}
\end{equation}

The cost of delivering downstream non-video traffic using hot potato routing is: 
\begin{equation}
\begin{aligned}
C^{ISP}_{S,U,non-video}= c^b V_d ED^{hot}_{down}(M)
\end{aligned}
\end{equation}
where $c^b$ is the cost per unit distance and per unit volume in the backbone network, and $ED^{hot}_{down}(M)$ is the average distance on the ISP’s backbone network of downstream non-video traffic with hot potato routing, when interconnecting at $M=12$ IXPs.

The cost of delivering downstream video traffic is the sum of the costs of delivering localized and non-localized video traffic:
\begin{equation}\label{isp_video}
\begin{aligned}
C^{ISP}_{S,U,video} = c^b V_v \Bigl[x ED^{cold}_{down}(M)+(1-x) ED^{hot}_{down}(M)\Bigl]
\end{aligned}
\end{equation}
The first term is the ISP's backbone cost for localized video traffic, which the transit provider delivers using cold potato routing. The second term is the ISP's backbone cost for non-localized video traffic, which the transit provider delivers using hot potato routing.  

The cost of delivering upstream traffic using hot potato routing is: 
\begin{equation}\label{eq:ISP_TP_UP}
\begin{aligned}
C^{ISP}_{U,S}
&= c^b V_u ED^{hot}_{up}(M) \\
&= c^b V_u ED^{cold}_{down}(M),
\end{aligned}
\end{equation}

Using the definition of the two traffic ratios $r$ and $r'$, and the fact that $ED^{cold}_{down}(M)=0$, equations (\ref{C_isp_eq})-(\ref{eq:ISP_TP_UP}) can be simplified as in Theorem \ref{theorem:c_isp}:
\begin{theorem}\label{theorem:c_isp}
The traffic-sensitive backbone cost of the ISP when peering with the transit provider is:    
\begin{equation}\label{c_isp}
\begin{aligned}
C^{ISP} 
&=c^b\Bigl[V_d+V_v(1-x)\Bigl]ED^{hot}_{down}(M)\\
&= c^b V_u \Bigl[r+r'(1-x)\Bigl]ED^{hot}_{down}(M) 
\end{aligned}
\end{equation}
\end{theorem}

A portion of the ISP's backbone cost is caused by the need to transport downstream non-video traffic over the ISP's backbone, as measured by the volume $V_d$ of such traffic.  Another portion of the ISP's backbone cost is caused by the need to transport downstream non-localized video traffic over the ISP's backbone, as measured by the volume $V_v(1-x)$ of such traffic.

\begin{figure}
\centering
\subfigure[$r=0.25$]{
\includegraphics[width=0.33\textwidth]{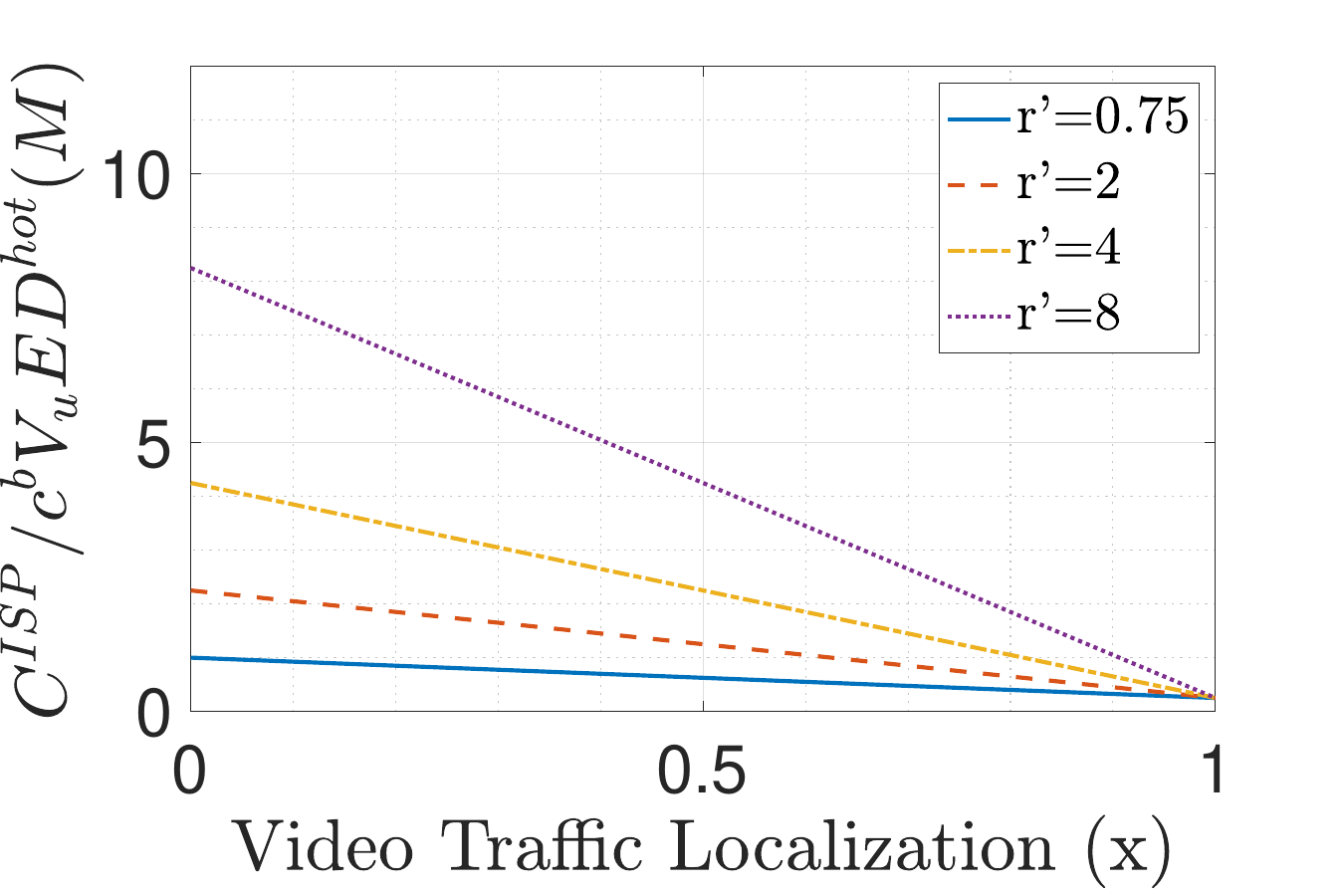}
}
\subfigure[$r=1$]{
\includegraphics[width=0.33\textwidth]{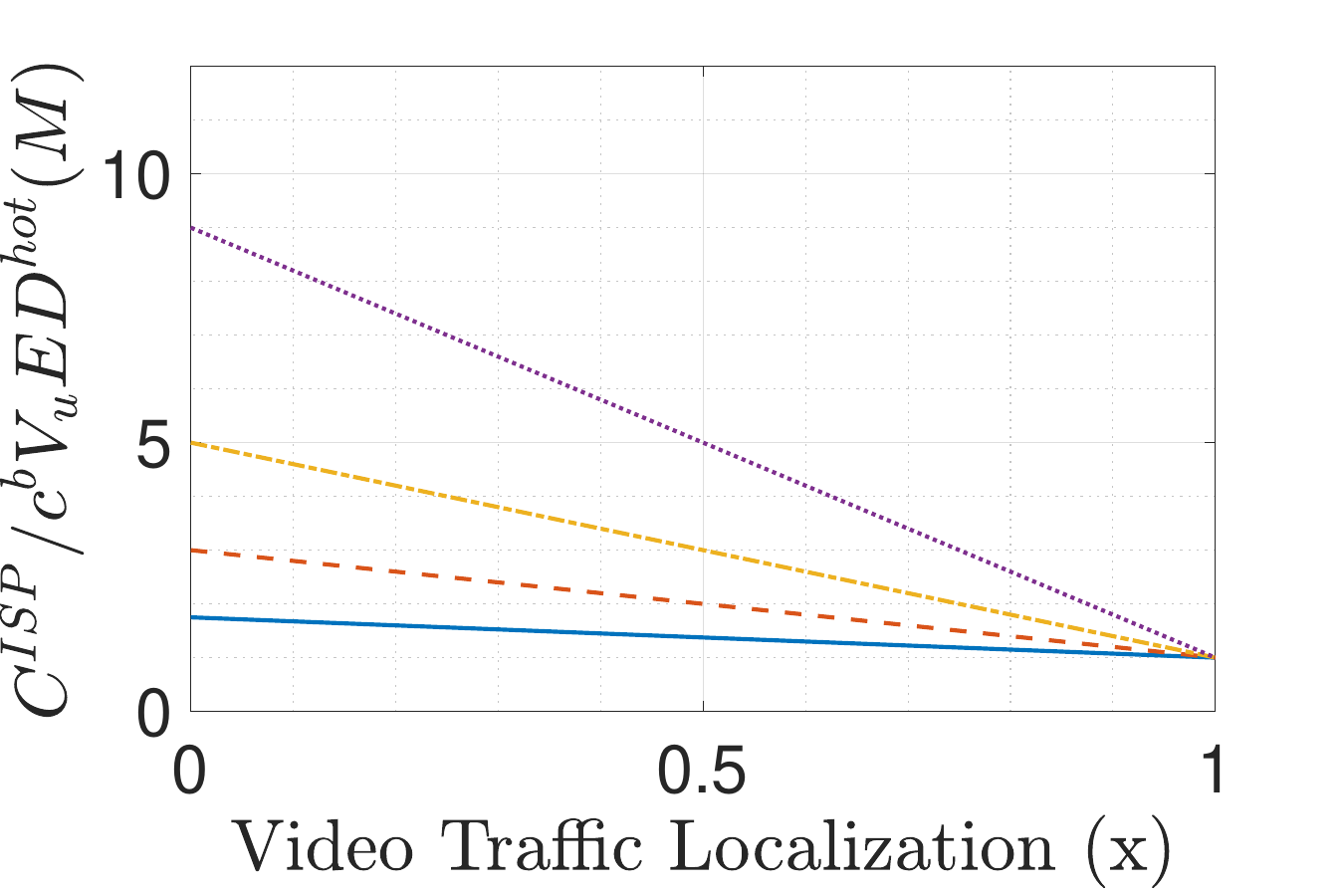}
}
\subfigure[$r=4$]{
\includegraphics[width=0.33\textwidth]{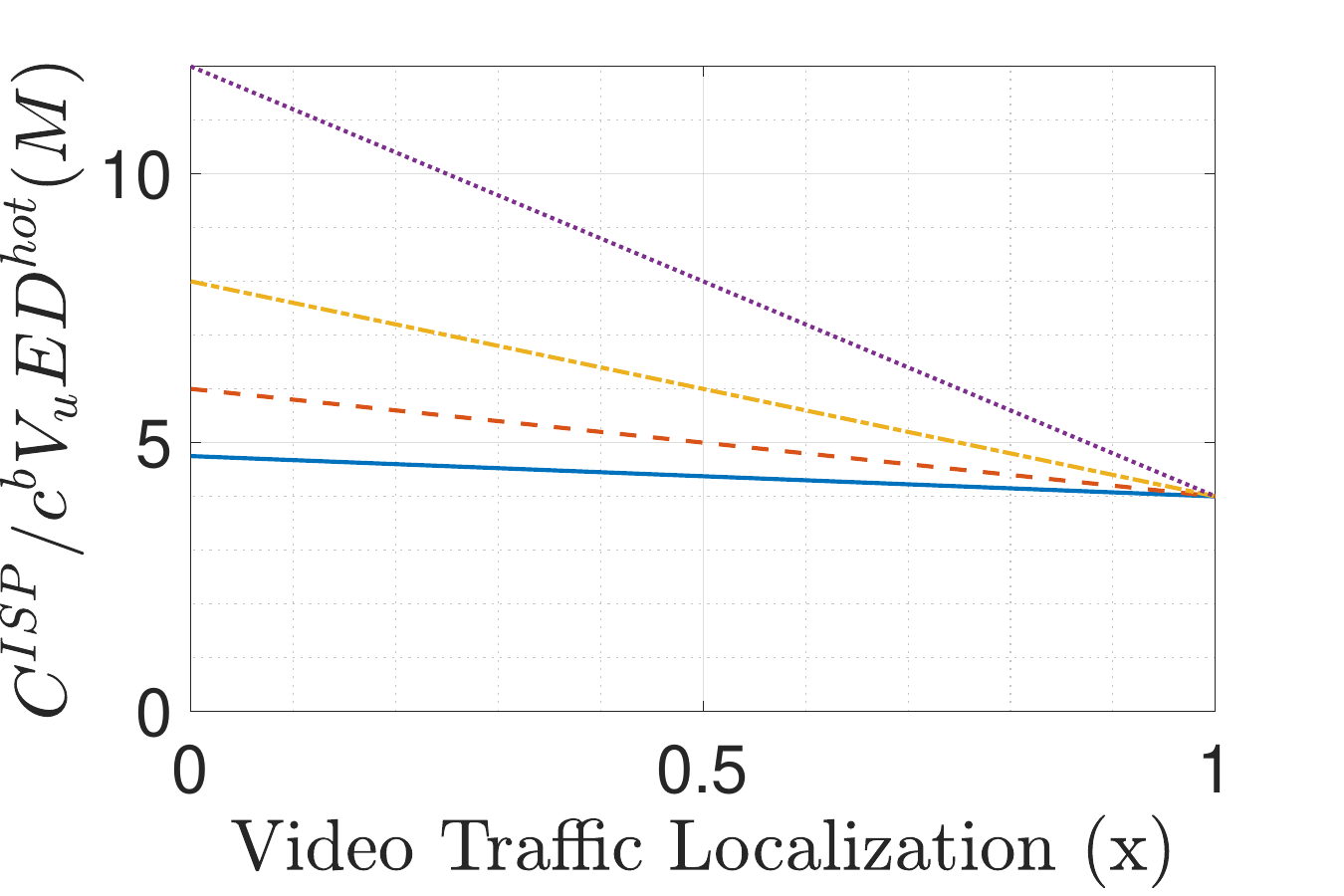}
}
\caption{ISP Backbone Cost}
\label{ISP_Cost}
\end{figure}

Figure \ref{ISP_Cost} illustrates the normalized ISP backbone cost for different traffic ratios and video traffic localization. For fixed traffic ratios ($r$ and $r'$), the normalized ISP backbone cost is decreasing with the amount of video traffic localization ($x$), because increasing localization results in the transit provider carrying more of the video traffic on its backbone network and handing it off to the ISP at an IXP closer to end users.  

For fixed non-video traffic ratio ($r$) and fixed video traffic localization ($x$), the normalized ISP backbone cost is increasing with the video traffic ratio ($r'$), because as video traffic increases, the ISP needs to carry more video traffic on its backbone. However, the amount of the increase in ISP backbone cost due to increased video traffic lessens as video traffic localization increases, because the ISP does not need to carry as much video traffic over long distances. 

Finally, for fixed video traffic ratio ($r'$) and fixed video traffic localization ($x$), the normalized ISP backbone cost is increasing with the non-video traffic ratio ($r$), because the ISP needs to carry more non-video downstream traffic on its backbone.

\subsection{Transit Provider Cost}

We now turn to the effect of routing policies, traffic ratios, and traffic localization on the traffic-sensitive backbone cost of the transit provider.  

Note that downstream traffic from the point of view of the transit provider (traffic entering the transit provider's network) is equal to the upstream traffic from the point of view of the ISP (traffic leaving the ISP's network) ($V_u$).  Similarly, the transit provider non-video upstream traffic is equal to ISP non-video downstream traffic ($V_d$), and the transit provider video upstream traffic is equal to ISP video downstream traffic ($V_v$).

We denote the transit provider's traffic-sensitive backbone cost by $C^{TP}$, and partition it into the transit provider's downstream cost for delivering ISP upstream traffic, which flows from the user $U$ to the source $S$, (denoted by $C^{TP}_{U,S}$), its upstream cost for delivering ISP downstream non-video traffic, which flows from the source $S$ to the user $U$, (denoted by $C^{TP}_{S,U,non-video}$), and its upstream cost for delivering ISP downstream video traffic, which flows from the source $S$ to the user $U$, (denoted by $C^{TP}_{S,U,video}$):
\begin{equation}\label{cost_tp}
\begin{aligned}
C^{TP}=C^{TP}_{U,S}+ C^{TP}_{S,U,non-video}+C^{TP}_{S,U,video} 
\end{aligned}
\end{equation}

The downstream cost to the transit provider for delivering ISP upstream traffic using hot potato routing is: 
\begin{equation}
\begin{aligned}
C^{TP}_{U,S}= c^b V_u ED^{hot}_{down}(M)
\end{aligned}
\end{equation}

The upstream cost to the transit provider for delivering ISP downstream non-video traffic using hot potato routing is: 
\begin{equation}
\begin{aligned}
C^{TP}_{S,U,non-video}
&= c^b V_d ED^{hot}_{up}(M)\\
&= c^b V_d ED^{cold}_{down}(M)
\end{aligned}
\end{equation}

The upstream cost to the transit provider for delivering ISP downstream video traffic is the sum of the costs of delivering localized and non-localized video traffic:
\begin{equation}\label{tp_video}
\begin{aligned}
C^{TP}_{S,U,video} 
&= c^b V_v \Bigl[x ED^{cold}_{up}(M)+(1-x) ED^{hot}_{up}(M)\Bigl] \\
&= c^b V_v \Bigl[x ED^{hot}_{down}(M)+(1-x) ED^{cold}_{down}(M)\Bigl]
\end{aligned}
\end{equation}
The first term is the transit provider's backbone cost for localized video traffic, which the transit provider delivers using cold potato routing. The second term is the transit provider's backbone cost for non-localized video traffic, which the transit provider delivers using hot potato routing.  

Using the definition of the two traffic ratios $r$ and $r'$, and the fact that $ED^{cold}_{down}(M)=0$, equations (\ref{cost_tp})-(\ref{tp_video}) can be simplified as in Theorem \ref{theorem:c_tp}:
\begin{theorem}\label{theorem:c_tp}
The traffic-sensitive backbone cost of the transit provider when peering with the ISP is: 
\begin{equation}\label{c_tp}
\begin{aligned}
C^{TP} 
&=c^b(V_u+V_v x)ED^{hot}_{down}(M)\\
&= c^b V_u (1+r' x)ED^{hot}_{down}(M) 
\end{aligned}
\end{equation}
\end{theorem}

A portion of the transit provider's backbone cost is caused by the need to transport the ISP's upstream traffic over the transit provider's backbone, as measured by the volume $V_u$ of such traffic.  Another portion of the transit provider's backbone cost is caused by the need to transport ISP downstream localized video traffic over the transit provider's backbone, as measured by the volume $V_v x$ of such traffic.

\begin{figure}
\centering
\includegraphics[width=\columnwidth]{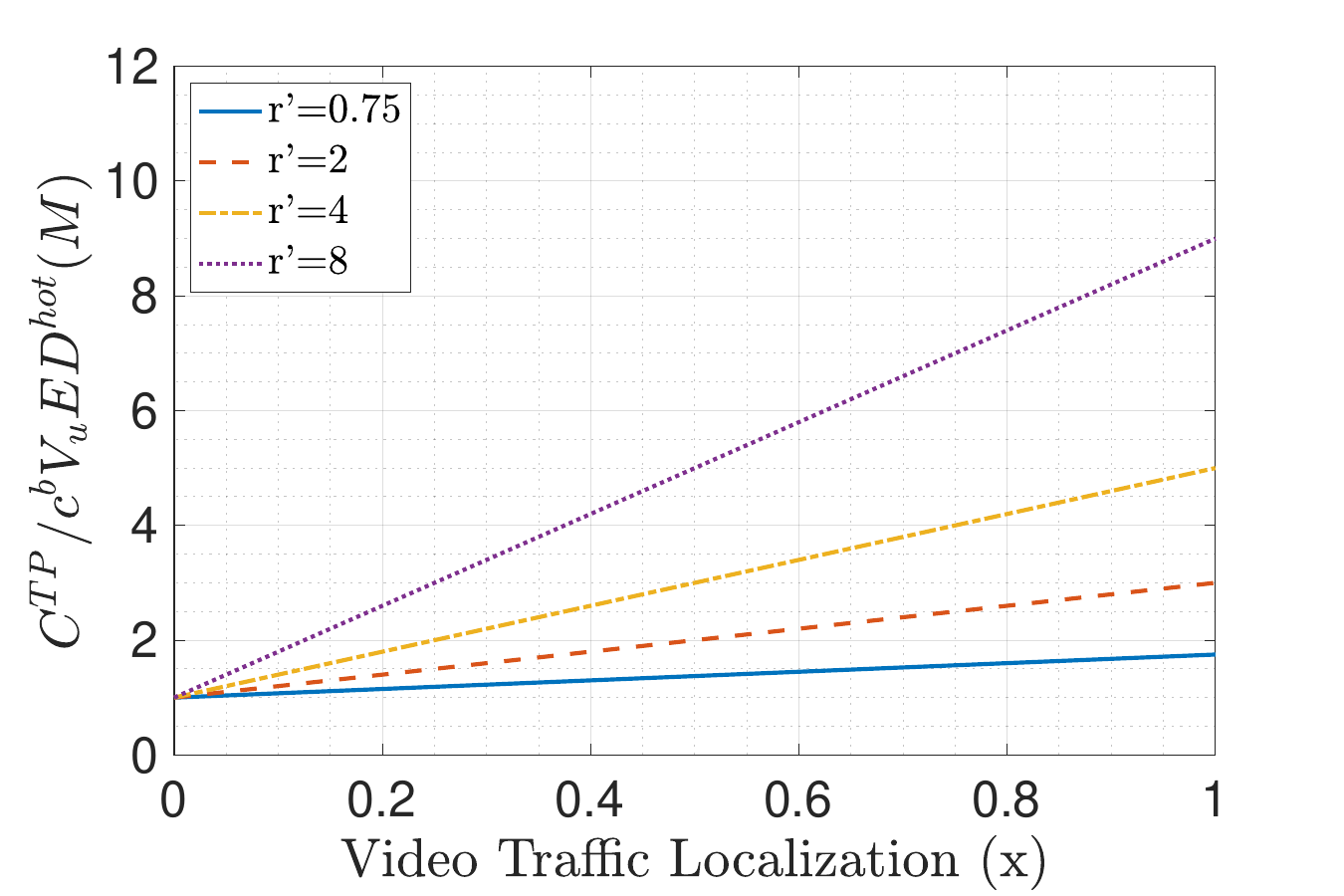}
\caption{Transit Provider Backbone Cost}
\label{TP_Cost}
\end{figure}

Figure \ref{TP_Cost} illustrates the normalized transit provider backbone cost for different video traffic ratios and video traffic localization. For a fixed video traffic ratio ($r'$), the normalized transit provider backbone cost is increasing with the amount of video traffic localization ($x$), because increasing localization results in the transit provider carrying more of the video traffic on its backbone network and handing it off to the ISP at an IXP closer to end users.  

For fixed video traffic localization ($x$), the normalized transit provider backbone cost is increasing with the video traffic ratio ($r'$), because as video traffic increases, the transit provider needs to carry more video traffic on its backbone. In addition, the amount of the increase in transit provider backbone cost due to increased video traffic increases as video traffic localization increases, because the transit provider needs to carry more video traffic over long distances.

The cost of the transit provider is normalized to the upstream traffic, so it remains constant with changes in the non-video traffic ratio ($r$), as the transit provider is only responsible for carrying the ISP upstream traffic on its backbone. Therefore, the figure is independent of $r$.

\subsection{Summary}

In this section, we analyzed the effect of routing policies, traffic ratios, and traffic localization on the traffic-sensitive backbone costs incurred by an ISP and a transit provider.  We found that as video traffic localization increases, the transit provider carries an increasing amount of the video traffic across its network and hands it off closer to end-users, leading to a decrease in the ISP's cost and an increase in the transit provider's cost.  Moreover, for a fixed percentage of video traffic localization, as the volume of video traffic increases, the costs incurred by the ISP and the transit provider both increase, since they both need to transport more video traffic on their backbones.  The increase in cost is more pronounced for the ISP when the traffic localization is low, due to the longer distances that the ISP must transport the traffic on its backbone.  Similarly, the increase in cost is more pronounced for the transit provider when the traffic localization is high, because the transit provider must carry the traffic over longer distances on its backbone. In addition, the impact of the imbalance in non-video traffic between the ISP and the transit provider also affects the ISP's cost share; as the ratio of non-video downstream traffic to upstream traffic increases, the ISP's cost share increases.

\section{Peering Between a Transit Provider and an ISP} \label{sec:Indirect}

In the previous section, we analyzed the traffic-sensitive backbone costs of the ISP and the transit provider. In this section, we determine the fair peering fee between the ISP and the transit provider. When analyzing peering between two ISPs, we defined \textit{fair} as the peering fee that equalized the net cost to each ISP. Here, when analyzing peering between an ISP and a transit provider, again we set the peering fee to equalize the net cost to each.

We consider three different scenarios based on how the transit provider delivers the traffic. First, we examine the case where the transit provider delivers the traffic with hot potato routing. Second, we explore the scenario in which the transit provider delivers part of the video traffic with cold potato routing, which localizes the traffic on the ISP's network. Finally, we consider the case where the transit provider uses a CDN to deliver part of the video traffic instead of delivering it with cold potato routing. Through these scenarios, we investigate the impact of different methods of delivering traffic on the cost-sharing arrangements between the ISP and the transit provider.

\subsection{Fair Peering Fee with Hot Potato Routing}

Recall that we denote the volume of non-video downstream traffic by $V_d$, the volume of video downstream traffic by $V_v$, and the volume of upstream traffic by $V_u$. We also define two traffic ratios: $r=\frac{V_d}{V_u}$, the ratio of downstream non-video traffic to upstream traffic, and $r'=\frac{V_v}{V_u}$, the ratio of downstream video traffic to upstream traffic. In this section, we assume the transit provider delivers all downstream traffic using hot potato routing. 

The ISP's traffic-sensitive backbone cost is determined solely by the locations of the IXPs at which traffic is exchanged, the routing of the traffic, and the volume of the traffic.  Thus, it follows that the fair peering fee between the ISP and the transit provider is still determined by equalizing their net costs, as was done in Theorem \ref{theorem:p_isp_isp} for the case of two peering ISPs, except that we must now account for the added video traffic:
\begin{equation}\label{eq:26}
P^{TP,ISP}=\frac{1}{2} c^b (V_d+V_v-V_u) ED^{hot}_{down}(M)
\end{equation}

The added video traffic results in a higher fair peering fee than would be the case of peering between two ISPs that did not include the exchange of this video traffic.

\subsection{Fair Peering Fee with Cold Potato Routing}

Transit providers that sell transit services to content providers often promise the content provider that they will deliver the traffic to the terminating ISP using cold potato routing. The use of cold potato routing allows the transit provider to exercise greater control over the management of this traffic and thereby potentially improve its Quality of Service (QoS). Most of this content consists of video, and in the remainder of the paper, we use the term \textit{video} to refer to it.

We consider the scenario in which the transit provider delivers part of the video traffic with cold potato routing, which localizes the traffic on the ISP's network.  Specifically, we assume that a proportion $x$ of the video traffic transmitted to the ISP's users within each access network is delivered from the transit provider using cold potato routing. We assume that the remaining proportion $1-x$ of the video traffic transmitted to the ISP's users within each access network is delivered using hot potato routing, and that the source of this video traffic is independent of the location of the end user. We also assume that the transit provider delivers the non-video traffic using hot potato routing.

We focus on the cost-sharing framework between the ISP and the transit provider. Our objective is to establish a system that ensures both parties incur the same costs for transmitting data over their backbones. To achieve this goal, we use the analysis of the cost structures of ISPs and transit providers that we provided in the previous section, and calculate the payment required to ensure that the costs are fairly split between the two parties. The peering fee that equalizes net costs is given by:
\begin{equation}\label{eq:ind_pay}
\begin{aligned}
P^{TP,ISP}=\frac{1}{2} (C^{ISP}-C^{TP})
\end{aligned}
\end{equation}
If the ISP's traffic-sensitive backbone costs exceed those of the transit provider, then the fair peering fee is positive. If the transit provider's traffic-sensitive backbone costs exceed those of the ISP, then the fair peering fee is negative. By using Theorems \ref{theorem:c_isp} and \ref{theorem:c_tp}, the fair peering fee can be expressed as:
\begin{equation}\label{eq:28}
\begin{aligned}
P^{TP,ISP}=\frac{1}{2} c^b V_u \Bigl[\Bigl(r-1\Bigl)+\Bigl(r'(1-x)-r'x\Bigl)\Bigl] ED^{hot}_{down}(M) 
\end{aligned}
\end{equation}
The term $\frac{1}{2} c^b V_u (r-1) ED^{hot}_{down}(M)$ is the fair peering fee resulting from any imbalance in the non-video traffic between the ISP and transit provider, similar to the case in which two ISPs peer.  The term $\frac{1}{2} c^b V_u r'(1-x) ED^{hot}_{down}(M)$ is the ISP's traffic-sensitive backbone cost incurred by non-localized video traffic.  The term $-\frac{1}{2} c^b V_u r'x ED^{hot}_{down}(M)$ is the transit provider's traffic-sensitive backbone cost incurred by localized video traffic. The terms can be combined to give:
\begin{theorem}\label{theorem:P_ind}
The fair peering fee between the transit provider and the ISP is:
\begin{equation}\label{P_ind}
\begin{aligned}
P^{TP,ISP}=c^b V_u \Bigl[ \frac{1}{2}(r-1)+r'(0.5-x) \Bigl] ED^{hot}_{down}(M)
\end{aligned}
\end{equation}
\end{theorem}

\begin{figure}
\centering
\subfigure[$r=0.25$]{
\includegraphics[width=0.33\textwidth]{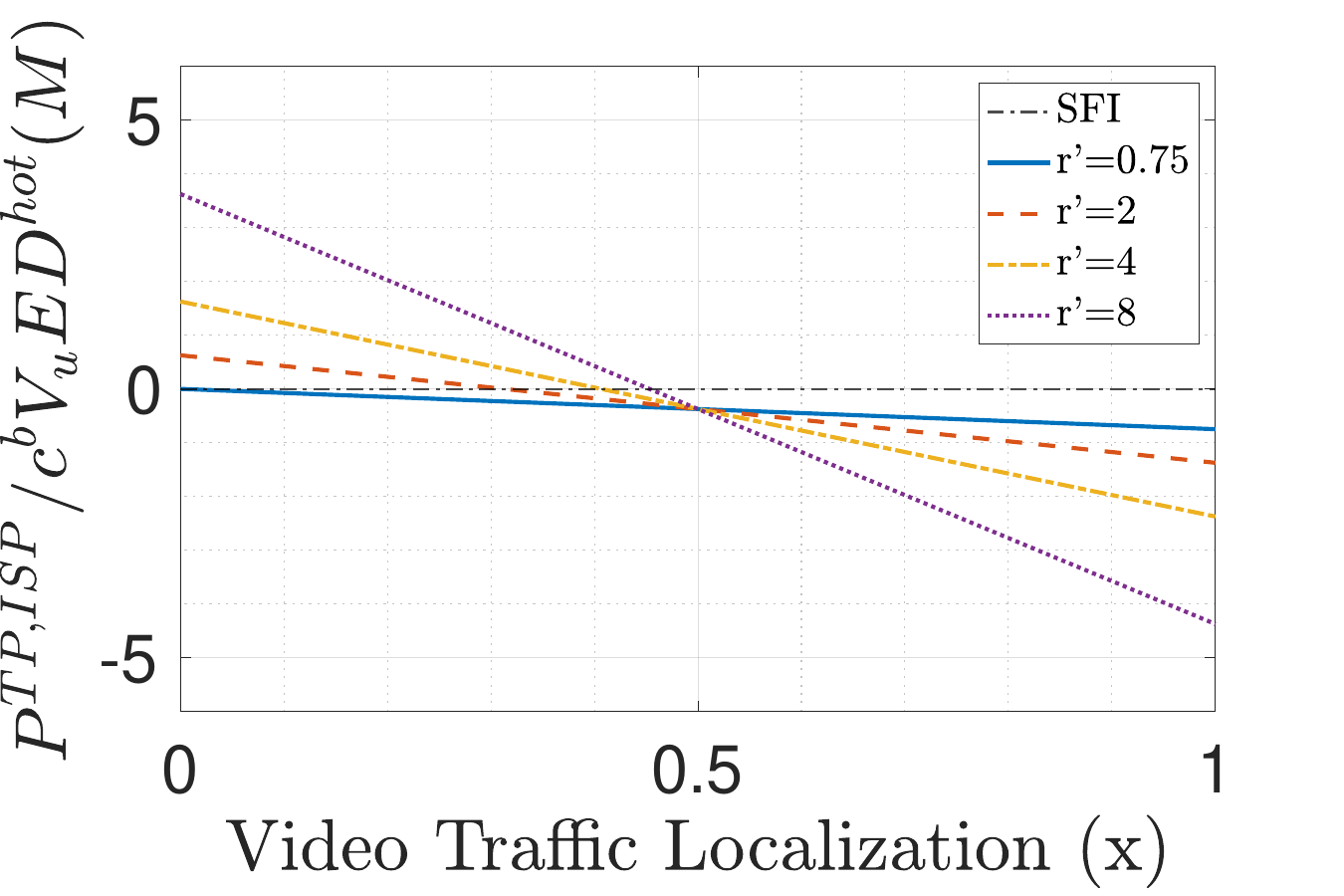}
}
\subfigure[$r=1$]{
\includegraphics[width=0.33\textwidth]{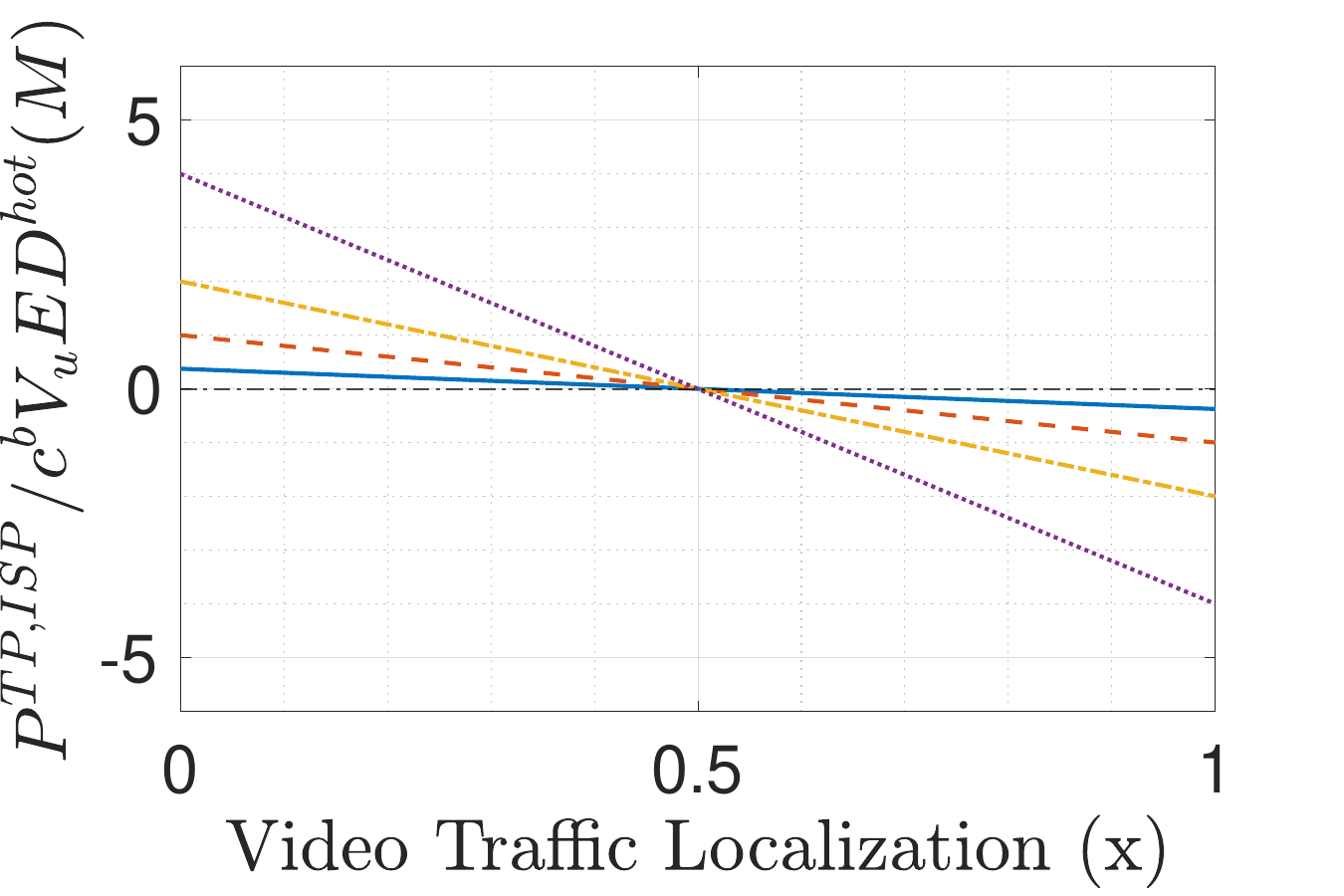}
}
\subfigure[$r=4$]{
\includegraphics[width=0.33\textwidth]{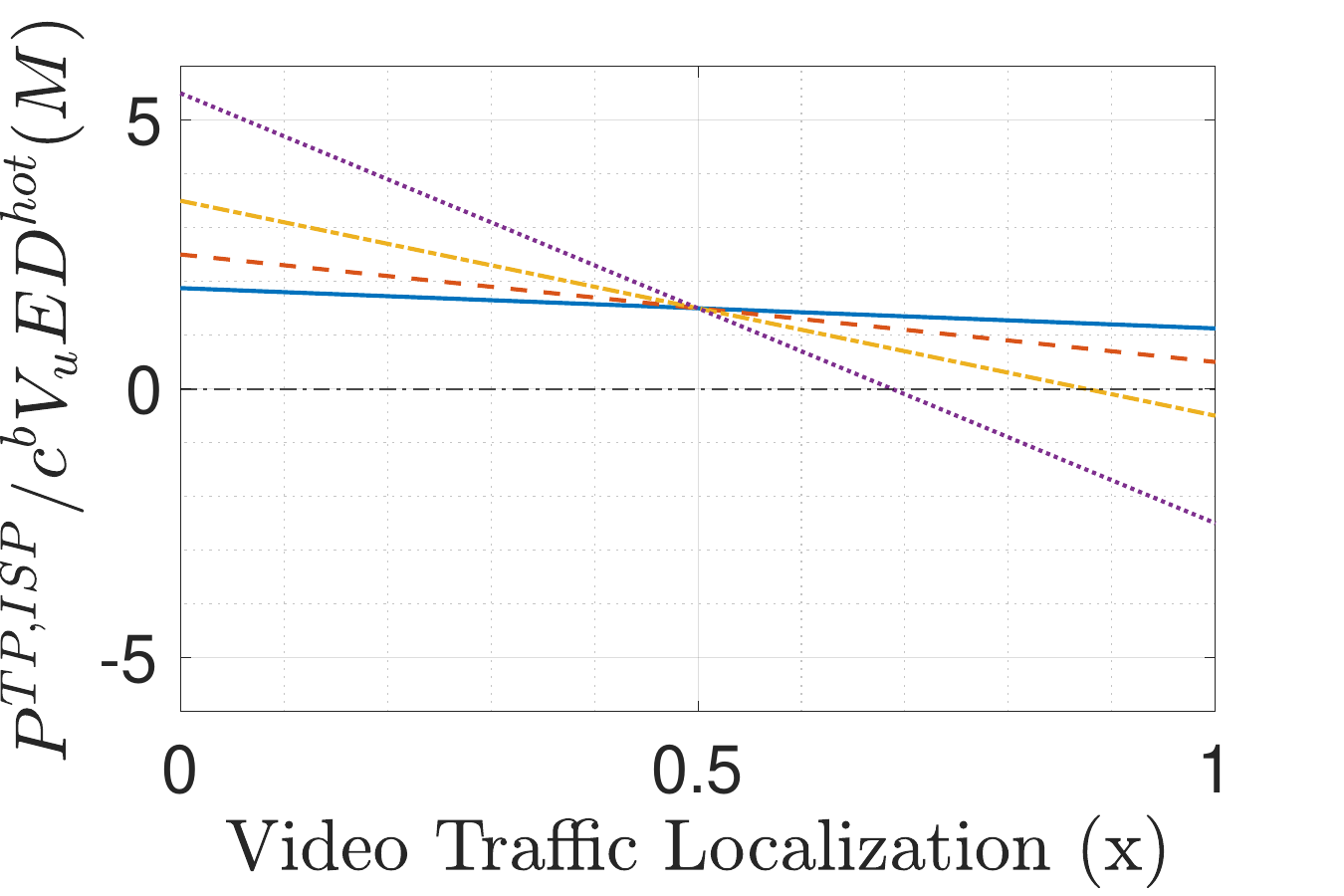}
}
\caption{Fair Peering Fee Between a Transit Provider and an ISP}
\label{Ind_Pay}
\end{figure}

Figure \ref{Ind_Pay} illustrates the normalized fair peering fee for different traffic ratios and video traffic localization. As the transit provider localizes an increasing percentage of video traffic, the fair peering fee decreases, reflecting the transit provider's increased cost and the ISP's decreased cost. Indeed, at high enough percentages of video traffic localization, the fair peering fee becomes negative, meaning that equalizing net cost requires the ISP to pay the transit provider (not vice versa). The slope at which the fair peering fee decreases with increasing video traffic localization becomes steeper at higher volumes of video traffic (i.e., higher video traffic ratios $r'$), reflecting a greater sensitivity of costs to the volume of video traffic. When video traffic localization is less than $50\%$, the fair peering fee increases with the volume of video traffic, since the transit provider sends more video traffic to the ISP without contributing much to its transportation cost. In contrast, when video traffic localization is more than $50\%$, the fair peering fee decreases as the amount of video traffic increases, since the transit provider contributes more to the cost of transporting this video across the backbone than does the ISP. Finally, for a fixed percentage of video traffic localization, the fair peering fee increases with the non-video traffic ratio $r$, reflecting the ISP's increased cost.

\subsection{Settlement-Free Peering}

Recall that when two ISPs peer, the fair peering fee is zero (i.e., settlement-free peering) when the traffic ratio $r=1$. In contrast, when a transit provider peers with an ISP, this is no longer the case. Now, absent any localization of video traffic, the fair peering fee is zero only if the combined traffic ratio $r+r'=1$. If the transit provider carries a substantial amount of video traffic, this is likely to result in a positive fair peering fee. However, the transit provider can reduce the fair peering fee by localizing a portion of the video traffic. Using Theorems \ref{theorem:c_isp} and \ref{theorem:c_tp}, the percentage of video traffic localization that equalizes the transit provider's and ISP's backbone costs is given by:
\begin{equation}
\begin{aligned}
c^b V_u \Bigl[r+r'(1-x)\Bigl]ED^{hot}_{down}(M) = c^b V_u \Bigl[1+r' x\Bigl]ED^{hot}_{down}(M)
\end{aligned}
\end{equation}

Solving for $x$ gives:
\begin{theorem}\label{theorem:x_ind}
The percentage of video traffic localization between a transit provider and an ISP required for a fair peering fee of zero is:
\begin{equation}
x=\frac{r+r'-1}{2r'}
\end{equation}
\end{theorem}

\begin{figure}
\centering
\includegraphics[width=\columnwidth]{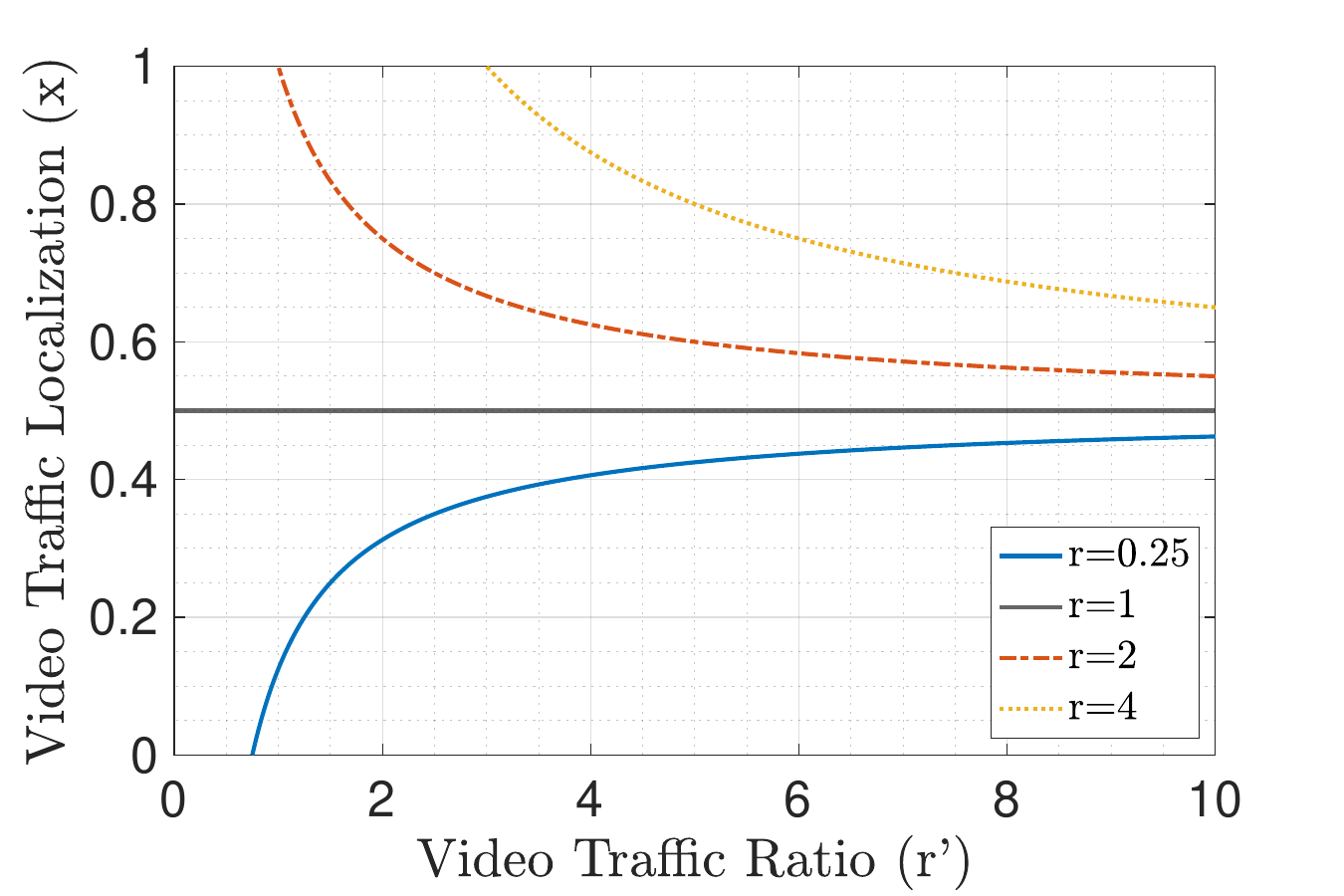}
\caption{Settlement-Free Peering Curve Between a Transit Provider and an ISP}
\label{fig:free_ind}
\end{figure}

Figure \ref{fig:free_ind} illustrates the amount of video traffic localization required for a fair peering fee of zero, as a function of the non-video traffic ratio $r$ and the volume of video traffic (reflected by the video traffic ratio $r'$). This settlement-free peering curve is determined by the relative costs incurred by the ISP and the transit provider.

When the non-video traffic ratio $r=1$, the volume of upstream and non-video downstream traffic is the same for both the ISP and the transit provider, resulting in both parties incurring equal non-video traffic-sensitive backbone costs. However, the addition of one-way video traffic sent via hot potato routing would impose extra costs on the ISP, which should be reimbursed through a positive peering fee. In order for the transit provider to achieve equal cost sharing with the ISP for the video traffic, it must localize 50\% of that traffic. If the transit provider localizes less than 50\% of the video traffic, the ISP incurs more backbone costs than does the transit provider, and consequently, compensation from the transit provider is warranted. Conversely, if the transit provider localizes more than 50\% of the video traffic, the transit provider incurs more backbone costs than does the ISP, and consequently compensation from the ISP is warranted. 

In contrast, when the non-video traffic ratio $r<1$, if there were no video traffic, the transit provider would incur more cost than the ISP, and consequently, the fair peering fee would be negative (i.e., the ISP should pay the transit provider). However, if there is a significant amount of video traffic sent via hot potato routing (e.g., $r' \approx 0.75$ and $x=0$), then the ISP cost to transport this video across its backbone can compensate for the unequal non-video traffic ratio (e.g., $r=0.25$), and result in settlement-free peering. As the volume of video traffic increases, in order to maintain equal net costs, the transit provider needs to start to localize some of the video traffic, i.e., the settlement-free peering curve is increasing with the volume of video traffic (as reflected by a higher video traffic ratio $r'$). The curve is concave and has an asymptote at $x=0.5$, because even at arbitrarily high video traffic volumes, 50\% localization is sufficient to entitle the transit provider for settlement-free peering. Indeed, it can be readily seen from Theorem \ref{theorem:x_ind} that $x \to 0.5$ as $r' \to \infty$.

Finally, when the non-video traffic ratio $r>1$, if there were no video traffic, the ISP would incur more cost than the transit provider, and consequently, the fair peering fee would be positive (i.e., the transit provider should pay the ISP). If there is video traffic but the transit provider transports it with less than 50\% localization, the cost burden on the ISP increases even more since the ISP incurs more cost to transmit this video than does the transit provider; hence, the transit provider should pay the ISP for both the high non-video traffic ratio and the imbalance in the cost of carrying video traffic. On the other hand, if the transit provider highly localizes video traffic, the transit provider may be eligible for settlement-free peering. For example, when $r=2$, if $r'=1$ and $x=1$, then the transit provider cost to transport this video across its backbone can compensate for the unequal non-video traffic ratio, and result in settlement-free peering. As the volume of video traffic increases, then in order to maintain equal net costs, the transit provider needs less localization, i.e., the settlement-free peering curve is decreasing with the volume of video traffic (as reflected by a higher video traffic ratio $r'$). The curve is convex and has an asymptote at $x=0.5$.

\subsection{Transit Provider Using CDNs}

In this section, we explore a scenario where the transit provider decides to implement Content Delivery Networks (CDNs) instead of carrying video traffic across its network using cold potato routing. 

Assume that the non-video traffic ratio $r$ and the video traffic ratio $r'$ are fixed, and that the transit provider decides to localize a proportion $x$ of the video traffic by transporting across its backbone using cold potato routing. Suppose that the ISP and the transit provider agree to the fair peering fee, as given in Theorem \ref{theorem:P_ind}. 

Now suppose that instead of localizing a proportion $x$ of the video traffic by transporting across its backbone using cold potato routing, the transit provider places this same video traffic on a CDN instead of delivering it using cold potato routing. Specifically, we assume that a proportion $x$ of the video traffic transmitted to the ISP's users within each access network is delivered from the transit provider at a CDN located at the IXP nearest to the end user. We assume that the remaining proportion $1-x$ of the video traffic transmitted to the ISP's users within each access network is delivered using hot potato routing, and that the source of this video traffic is independent of the location of the end user.

By using a CDN, the transit provider reduces its traffic-sensitive backbone cost by:
\begin{equation}
\Delta C^{TP} = c^b V_v x ED^{hot}_{down}(M)
\end{equation}
The result follows from Theorem \ref{theorem:c_tp}.

However, in order to build the CDN, the transit provider incurs a cost which we denote by $Cost_{tp}^{CIC}$. If $Cost_{tp}^{CIC} < \Delta C^{TP}$, the transit provider may be incentivized to build the CDN. However, the question arises whether building the CDN will affect the fair peering fee. Indeed, the ISP may assert that it should share in a portion of the cost savings by increasing the fair peering fee.

We reject the notion that building the CDN should affect the fair peering fee. If the fair peering fee is not affected, then the transit provider will make a decision based solely upon a comparison between the cost of implementing CDNs with the cost of carrying traffic using cold potato routing:
\begin{theorem}\label{theorem:delta_P_ind}
The transit provider can achieve cost savings by implementing CDNs at $M$ interconnection points with $x$ percent of localization, if the cost of implementation is lower than the potential savings:
\begin{equation}
Cost_{tp}^{CIC}<\Delta C^{TP}= c^b V_v x ED^{hot}_{down}(M)
\end{equation}
\end{theorem}

This comparison of the cost of servers versus transmission is a classical engineering tradeoff. In contrast, if the fair peering fee were to be increased in order to share some of the cost savings with the ISP, then it would reduce the incentive of the transit provider to build a CDN, resulting in an inefficient architecture. In addition, we note that without a change in the fair peering fee, the ISP's net costs are unchanged.

\subsection{Evaluation of Arguments}

Our results contradict the manner in which large ISPs often portray the situation. Large ISPs often assert that the fair peering fee is positive whenever the combined traffic ratio $r+r'>1$ regardless of the amount of localization. In contrast, we find that from Theorem \ref{theorem:x_ind} that when $r+r'>1$, the fair peering fee is positive if and only if $x < \frac{r+r'-1}{2r'}$. For example, if $r=1$, then a positive peering fee is only warranted if $x<0.5$. In addition, large ISPs often assert that the fair peering fee increases monotonically with $r+r'$ regardless of the amount of localization. We do find that when video traffic localization is low, the fair peering fee increases monotonically with $r+r'$. However, for high levels of localization, the fair peering fee decreases with $r+r'$, since the transit provider incurs most of the backbone transportation cost.

Our results also partially contradict the manner in which transit providers often portray the situation. Transit providers often assert that they should be entitled to settlement-free peering if they provide sufficient localization of traffic. Although we find that this is true when the non-video traffic ratio $r<1$, for high non-video traffic ratios (e.g., $r=4$), even 100\% localization of video traffic may not be sufficient unless the transit provider also localizes non-video traffic. 

\subsection{Summary}

In this section, we analyzed the fair fee for peering between a transit provider and an ISP. If the transit provider uses hot potato routing for all traffic, then the fair peering fee is given by (\ref{eq:26}), which states that it is a function of the imbalance between all download traffic ($V_d + V_v$) and upload traffic ($V_u$). 

If the transit provider uses cold potato routing for a proportion $x$ of the video traffic, then the fair peering fee is given by Theorem \ref{theorem:P_ind}. The fair peering fee increases with the non-video traffic ratio $r$ and decreases with the proportion $x$. It also decreases more rapidly with $x$ for higher video traffic volumes. The transit provider should pay the ISP for peering if it doesn't localize a sufficient portion of the video traffic. The fair peering fee may be positive and substantial if there is a high volume of video traffic with low localization.

The amount of localization that equalizes net costs is given by Theorem \ref{theorem:x_ind}. Settlement-free peering is appropriate when the transit provider localizes a sufficient proportion of video traffic. The required proportion is less than 0.5 when the non-video traffic ratio $r<1$, equal to 0.5 when the non-video traffic ratio $r=1$, and greater than 0.5 when the non-video traffic ratio $r>1$. 

Finally, we argue that the fair peering fee should be unchanged if the transit provider uses a CDN to localize traffic instead of delivering it using cold potato routing. If so, a CDN will result in cost savings if the cost of building it is less than the cost of carrying traffic across the transit provider's backbone.

\section{Peering Between a Content Provider and an ISP}\label{sec:Direct}

We now turn to peering between a content provider and an ISP. We compare such direct interconnection with the ISP with the indirect interconnection considered in the previous section, in which a content provider sends video traffic through a transit provider to the ISP. We focus on the impact of elements of peering policies, including the number of interconnection points and video traffic localization, on the fair peering fee. We also determine the conditions under which a content provider should be eligible for settlement-free peering.

There are three key differences that may make the fair peering fee between a content provider and an ISP different than that between a transit provider and an ISP. The first difference is that the content provider may choose to interconnect at a lower number of IXPs, which could potentially increase the cost for the ISP. The second difference is that the content provider may choose a different localization strategy compared to the transit provider, which could also potentially impact the cost of the ISP. The third difference is that there is only video traffic between the content provider and the ISP.

In addressing ISP-content provider peering, our analysis remains neutral to the content provider's network topology. We focus on calculating a fair peering fee based only on the ISP's costs, which are influenced by interconnection points and traffic localization, not by whether the content provider uses its own backbone or a CDN. This ensures that our determination of a fair peering fee is consistent regardless of the content provider's infrastructure.

\subsection{ISP's Backbone Cost}

In this subsection, we consider the traffic-sensitive backbone costs of the content provider and of the ISP. In the following subsection, we determine the fair peering fee. 

We assume that both the ISP and the content provider have agreed to peer with each other, however, peering between a content provider and an ISP differs from peering between a transit provider and an ISP for three reasons. First, we assume that the content provider has deployed content servers at $N$ major interconnection points, and that $N$ may be less than the number of interconnection points at which the transit provider and the ISP agree to peer ($M$). This could increase the ISP's backbone cost. 

Second, the content provider may localize a different proportion of video traffic than does the transit provider, which may impact the ISP's backbone cost. A higher degree of localization may reduce the backbone costs for the ISP. We assume that a proportion $x^d$ of the video transmitted to the ISP's users within each access network is delivered from the content provider at a server located at the IXP nearest to the end user among the IXPs at which they agree to peer. We assume that the remaining proportion $1-x^d$ of the video transmitted to the ISP's users within each access network is delivered using any content provider server, and that the location of this content provider's server is independent of the location of the end user.

Third, content providers carry only video traffic, which results in a higher ratio of downstream traffic (from the content provider to the ISP) to upstream traffic (from the ISP to the content provider) compared to transit providers. This is due to the fact that video traffic is almost entirely downstream. We denote the volume of video downstream traffic by $V_v$. 

We denote the ISP's traffic-sensitive backbone cost by $C^{ISP}_{cp,video}$. The cost of delivering downstream video traffic is the sum of the costs of delivering localized and non-localized video traffic:
\begin{theorem}\label{theorem:c_isp_cp}
The traffic-sensitive backbone cost of the ISP when peering with the content provider is:    
\begin{equation}
\begin{aligned}\label{c_isp_d}
C^{ISP}_{cp,video}=c^b V_v \Bigl[x^d ED^{cold}_{down}(N)+(1-x^d) ED^{hot}_{down}(N)\Bigl]
\end{aligned}
\end{equation}
\end{theorem}
The first term, $c^b V_v x^d ED^{cold}_{down}(N)$, is the ISP's backbone cost for localized video traffic, which accounts for the ISP's transport of a proportion $x^d$ of the video traffic from the IXP nearest to the end user among the IXPs at which they agree to peer. The second term, $c^b V_v (1-x^d) ED^{hot}_{down}(N)$, is the ISP's backbone cost for non-localized video traffic, which accounts for the ISP's transport of a proportion $1-x^d$ of the video traffic from any IXP where they have agreed to peer. 

\subsection{Fair Peering Fee}

We must first address the question of how to define \textit{fair} in the context of direct peering between a content provider and an ISP. Should we define \textit{fair} as the peering fee that equalizes the net costs of the content provider and the ISP, similar to our analysis above for peering between two ISPs or peering between a transit provider and an ISP? Or should we define \textit{fair} as the peering fee that results in the same ISP net costs for transporting the video traffic as in the case in which the video traffic is transported across a transit provider's network?

We believe that the appropriate definition of \textit{fair} is the latter one. If we were to attempt to equalize the net costs of the content provider and the ISP, we would have to account for the cost to the content provider of building its CDN. However, as we argued above in the case in which a transit provider deploys a CDN, the decision between building a CDN versus transporting video traffic across the backbone should be made on the basis of the cost of servers versus the transmission cost, not also on the peering fee. Direct peering between a content provider and an ISP is similar. Again, the fair peering fee should be determined solely by ensuring that the ISP's net costs are unaffected by the content provider's decision.  

Thus, we define the fair peering fee between a content provider and an ISP as the fee that results in the same ISP net costs for transporting the video traffic as in the case where the video traffic enters the ISP's network indirectly through a transit provider. Note, however, that this fair peering fee is different than that between a transit provider and an ISP for the three reasons discussed above. 

Denote the fair peering fee between the transit provider and the ISP that is related solely to video traffic (not related to upstream or non-video downstream traffic) by $P_{v}^{TP,ISP}$. Using Theorem \ref{theorem:P_ind}, we can determine $P_{v}^{TP,ISP}$ by considering only the video traffic component of the fair peering fee between the transit provider and the ISP ($P^{TP,ISP}$). It can be expressed as:
\begin{equation}\label{pt_tv_v}
\begin{aligned}
P_{v}^{TP,ISP} &= c^b V_v (0.5-x) ED^{hot}_{down}(M)
\end{aligned}
\end{equation}

Denote the fair peering fee for direct interconnection between a content provider and an ISP by $P^{CP,ISP}$. It is given by:
\begin{equation}\label{p_cp_d}
\begin{aligned}
P^{CP,ISP} = P_{v}^{TP,ISP} + (C^{ISP}_{cp,video} - C^{ISP}_{S,U,video})
\end{aligned}
\end{equation}
where $C^{ISP}_{cp,video}$ is the traffic-sensitive backbone cost of the ISP when peering with the content provider (given in Theorem \ref{theorem:c_isp_cp}), and $C^{ISP}_{S,U,video}$ is the traffic-sensitive backbone cost of the ISP for delivering video traffic when peering with a transit provider (given in (\ref{isp_video})).

The cost difference $(C^{ISP}_{cp,video} - C^{ISP}_{S,U,video})$ accounts for any changes to the ISP's cost resulting from any differences in traffic flows and localization when it peers with a content provider rather than with a transit provider.

Using (\ref{isp_video}), (\ref{pt_tv_v}) and Theorem \ref{theorem:c_isp_cp}, we can express the fair peering fee in (\ref{p_cp_d}) as:
\begin{equation}
\begin{aligned}
P^{CP,ISP}
&=c^b V_v \Bigl[x^d ED^{cold}_{down}(N)\\
&+(1-x^d) ED^{hot}_{down}(N)-0.5 ED^{hot}_{down}(M)\Bigl]
\end{aligned}
\end{equation}
Finally, we can rearrange the terms to separate the effects of the number of IXPs at which the content provider and the ISP peer from the effects of localization:
\begin{theorem}\label{theorem:P_d}
The fair peering fee between the content provider and the ISP is:
\begin{equation}\label{Pd}
\begin{aligned}
P^{CP,ISP}
&= {c^b V_v} \Bigl[\Bigl(0.5-x^d\Bigl)ED^{hot}_{down}(M) \\
&+ \Bigl(1-x^d\Bigl)\Bigl(ED^{hot}_{down}(N)-ED^{hot}_{down}(M)\Bigl)\\
&+x^d \Bigl(ED^{cold}_{down}(N)-ED^{cold}_{down}(M)\Bigl)\Bigl]
\end{aligned}
\end{equation}
\end{theorem}
The first term, $c^b V_v (0.5-x^d) ED^{hot}_{down}(M)$, represents the effect of video traffic localization on the fair peering fee. The second term, $c^b V_v (1-x^d) (ED^{hot}_{down}(N)-ED^{hot}_{down}(M))+x^d (ED^{cold}_{down}(N)-ED^{cold}_{down}(M))$, represents the effect of the number of IXPs at which they agree to peer on the fair peering fee.

\begin{figure}
\centering
\includegraphics[width=\columnwidth]{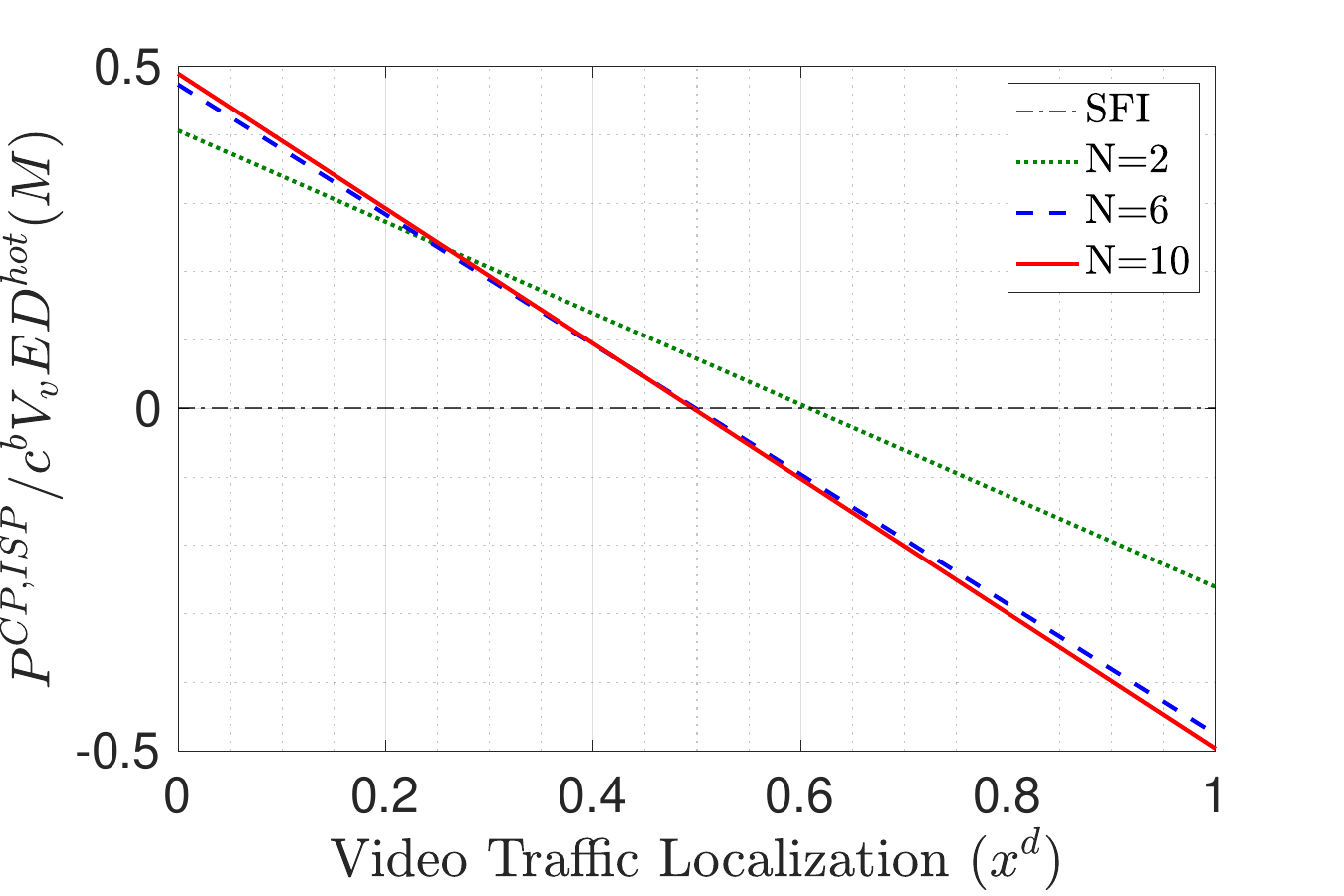}
\caption{Fair Peering Fee Between a Content Provider and an ISP}
\label{Dir_Pay}
\end{figure}

Figure \ref{Dir_Pay} illustrates the fair peering fee between the content provider and the ISP as a function of video traffic localization and the number of IXPs at which they agree to peer. At low amounts of localization, the fair peering fee is positive. However, as the content provider sends traffic with more localization, the fair peering fee decreases and at some point becomes negative (meaning that the ISP should pay the content provider). 

The fair peering fee also varies with the number of IXPs at which they agree to peer. When localization is very low, interconnecting at more IXPs is not beneficial to the ISP, because the ISP needs to carry the video traffic over longer distances in its backbone network since the peering IXP moves farther from the IXP nearest to the end user \cite{tprc_us}. Therefore, the fair peering fee increases slightly with $N$ at very low amounts of localization. However, for moderate to high localization, interconnecting at more IXPs is beneficial to the ISP, because the ISP's backbone cost decreases since the peering IXP moves closer to the IXP nearest to the end user \cite{tprc_us}. Therefore, the fair peering fee decreases with $N$ at moderate to high amounts of localization. 

\subsection{Settlement-Free Peering}

Finally, we wish to determine under what elements of peering policies (namely, the number of interconnection points and video traffic localization) the content provider should be eligible for settlement-free peering. By setting the fair peering fee to zero ($P^{CP,ISP}=0$) in Theorem \ref{theorem:P_d}, we can determine the number of IXPs and localization required for settlement-free peering:
\begin{theorem}\label{theorem:x_dir}
The percentage of video traffic localization between a content provider and an ISP required for a fair peering fee of zero is:
\begin{equation}
\begin{aligned}
x^d= \frac{ED^{hot}_{down}(N)-0.5 ED^{hot}_{down}(M)}{ED^{hot}_{down}(N)- ED^{cold}_{down}(N)} 
\end{aligned}
\end{equation}
\end{theorem}

\begin{figure}
\centering
\includegraphics[width=\columnwidth]{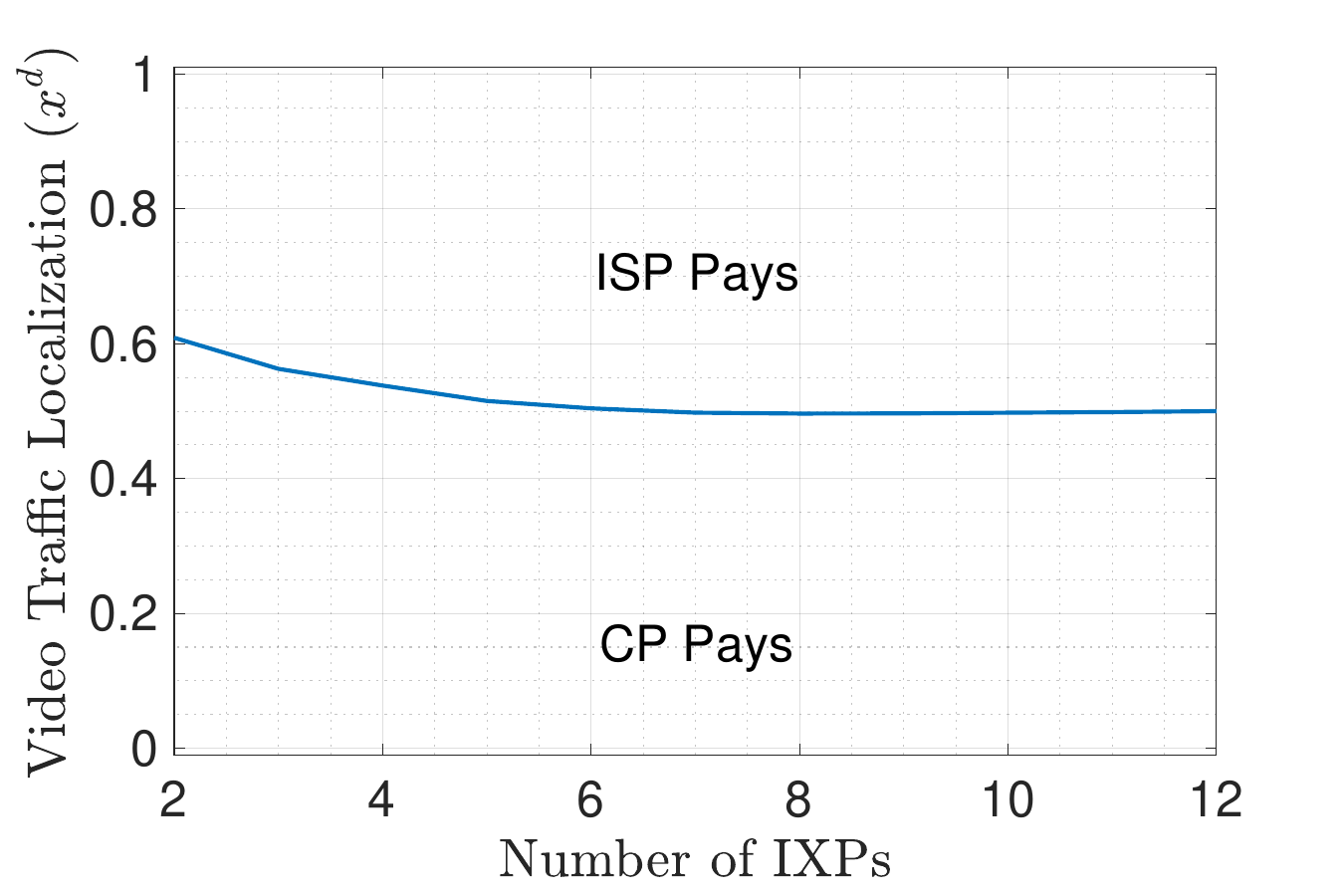}
\caption{Settlement-free Direct Peering Curve}
\label{fig:free_curve}
\end{figure}

Figure \ref{fig:free_curve} illustrates the settlement-free peering curve for direct interconnection between a content provider and an ISP, as a function of the amount of video traffic localization and the number of IXPs at which they peer. Recall from Figure \ref{fig:free_ind}, which illustrated the settlement-free peering curve for interconnection between a transit provider and an ISP, that 50\% localization is sufficient for settlement-free peering when the non-video traffic ratio $r=1$, but that different amounts of localization may be required for other non-video traffic ratios and it may depend on the amount of video traffic. For direct interconnection between a content provider and an ISP, the traffic ratio is now irrelevant because the net cost to the ISP of transporting the video traffic is solely a function of localization.\footnote{The required localization is solely dependent on the location of the interconnection points and the distribution of video traffic among the population. The other parameters of our model do not affect the settlement-free peering curve.} Indeed, if the content provider and the ISP agree to peer at all $N=12$ locations, then 50\% localization is sufficient to justify settlement-free peering.

As the number of interconnection points decreases from $N=12$, the content provider should send an increasing proportion of video traffic locally in order to be eligible for settlement-free peering. However, when $8 \leq N \leq 12$, there is little variation in ISP's backbone cost and thus little change in the amount of localization required for settlement-free peering. 

\subsection{Evaluation of Arguments}

Recall that large ISPs often argue that they should apply the same settlement-free peering requirements to peering ISPs, peering transit providers, and peering content providers. However, whereas when two ISPs peer we have shown that the fair peering fee is a function of the traffic ratio, when a content provider and an ISP peer we have shown that the fair peering fee is a function of the number of interconnection points and of localization. Hence, different settlement-free peering should apply to these two situations. 

Also, recall that large ISPs often argue that they should be compensated by large content providers regardless of the amount of video content localization. Again we disagree. Our results show that it is rational for an ISP to agree to settlement-free peering with a content provider that delivers a sufficient amount of video traffic locally.  

Similarly, recall that large content providers sometimes argue that they should be entitled to settlement-free peering solely because the ISP's customers have already paid the ISP to transport the traffic the content providers are sending. We believe this argument is too simplistic. We have shown that the fair peering fee should include consideration of the ISP's backbone transportation cost and thus of the number of interconnection points and of localization. A more nuanced argument by large content providers is that they should be eligible for settlement-free peering if they bring the content close to customers. We have shown that localization should indeed play a key role in determining eligibility for settlement-free peering.

\section{Conclusion}

In this paper, we derived the fair peering fee between two ISPs, the fair peering fee between a transit provider and an ISP which agree to peer with each other, and the fair peering fee between a content provider and an ISP which directly interconnect with each other. We analyzed the impact of routing policies, traffic ratios, and traffic localization on backbone costs for ISPs and transit providers. Our analysis showed that these factors play a crucial role in determining the costs and the fair fees for peering.

First, we analyzed the peering between two ISPs. We investigated how traffic ratios affect the costs and payments between the two ISPs. Our results show that symmetric ISPs would likely reach a settlement-free peering agreement. However, peering between ISPs with unequal traffic may require payment between the two networks, and the payment depends on the traffic ratio. 

Next, we examined how routing policies, traffic ratios, and traffic localization impact backbone costs for the ISP and the transit provider. We found that video traffic localization impacts ISP and transit provider backbone costs differently, with increasing video localization by the transit provider leading to decreased ISP cost but increased transit provider cost. Increasing volumes of video traffic increases costs for both, with the impact more pronounced for an ISP when traffic localization is low and for a transit provider when it's high. The ratio of non-video downstream to upstream traffic also affects an ISP's cost share; as the ratio of non-video downstream traffic to upstream traffic increases, an ISP's cost share increases.

We then examined the fair peering fee between a transit provider and an ISP. We define \textit{fair} as the peering fee that equalized the net backbone costs of the transit provider and the ISP. We found that the fair peering fee for a transit provider using hot potato routing depends on the downstream-upstream traffic ratio, while for a transit provider using cold potato routing for video traffic, it depends on the proportion of localized video traffic, the non-video traffic ratio, and the volume of video traffic. The fair peering fee increases with the non-video traffic ratio and decreases with the proportion of localized video traffic. It also decreases more rapidly with the proportion of localized video traffic for higher volumes of video traffic. A transit provider should pay an ISP for peering if it doesn't localize a sufficient proportion of the video traffic. The fair peering fee may be positive and substantial if there is a high volume of video traffic with low localization.

Finally, we examined the fair peering fee between a content provider and an ISP. Now, we define \textit{fair} as the peering fee that results in the same ISP net costs for transporting the video traffic as in the case in which the video traffic is transported across a transit provider's network. Our results indicate that settlement-free peering is solely dependent on the localization of video traffic and the number of interconnection points. We found that as the number of interconnection points decreases, the content provider should increase the proportion of locally sent video traffic to maintain eligibility for settlement-free peering.

In conclusion, we expect that an ISP should have different settlement-free peering requirements for content providers than for other ISPs. We also expect that the settlement-free peering requirements for content providers may include a specified minimum number of interconnection points and a specified minimum amount of traffic to be delivered locally. However, we certainly expect there to be \textit{no} traffic ratio requirements. 

In future work, we intend to determine the peering price that maximizes an ISP's profit, based on the costs analyzed here as well as on its power to earn a profit over these costs. (Earlier, we considered profit maximization in \cite{me_telecom}, but without a detailed cost model.) Our goal is to determine the range of peering prices from the cost-based peering price (at the low end) to the profit-maximizing peering price or the maximum willingness-to-pay (whichever is less).

\section*{Acknowledgment}
This material is based upon work supported by the National Science Foundation under Grant No. 1812426.

\bibliography{References}
\bibliographystyle{IEEEtran}

\begin{IEEEbiography}[{\includegraphics[width=1in,height=1.25in,clip,keepaspectratio]{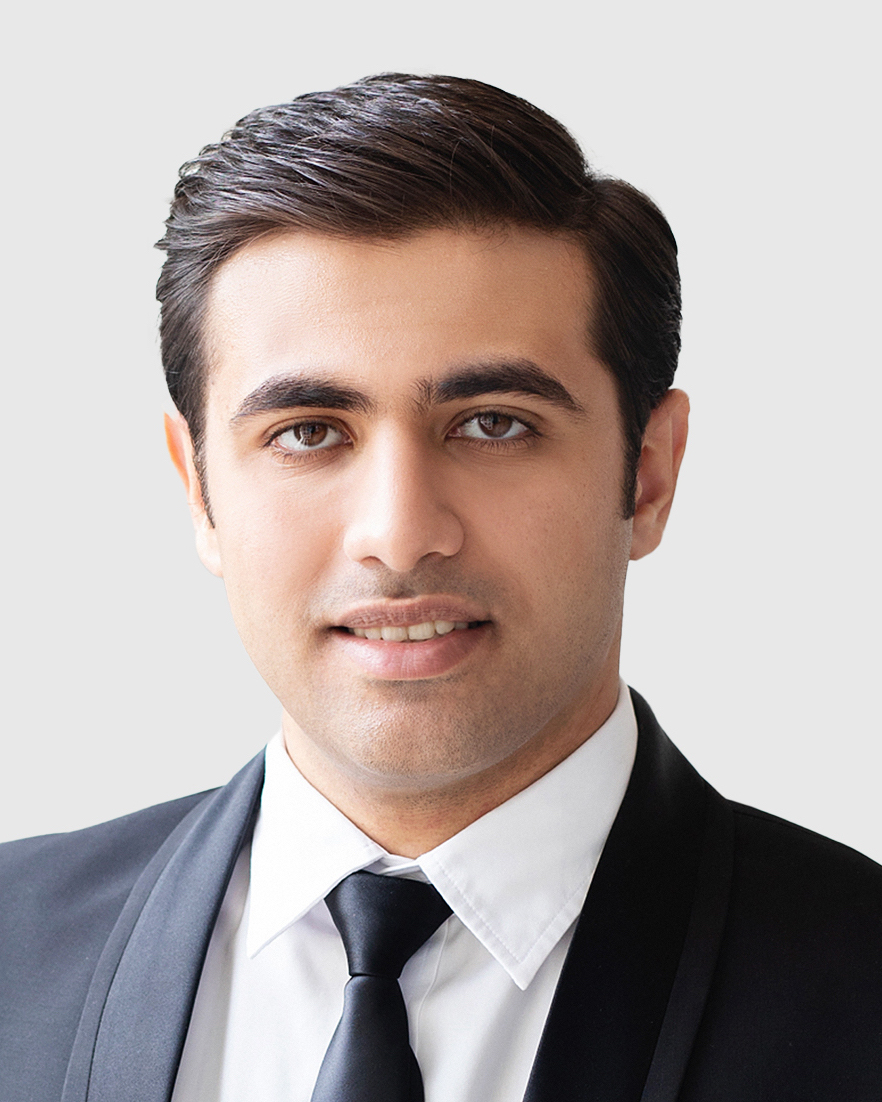}}]{Ali Nikkhah}
is a Ph.D. candidate in the Networked Systems program at the University of California, Irvine. He obtained his B.Sc. and M.Sc. degrees in Electrical Engineering from the Sharif University of Technology in 2018 and 2016, respectively. Currently, Ali is working on an NSF grant project that focuses on internet interconnection policy. His interests lie at the intersection of telecommunication policy and data science. In addition to his research, Ali is also an active member of the IEEE and has presented his work at several conferences.
\end{IEEEbiography}

\begin{IEEEbiography}[{\includegraphics[width=1in,height=1.25in,clip,keepaspectratio]{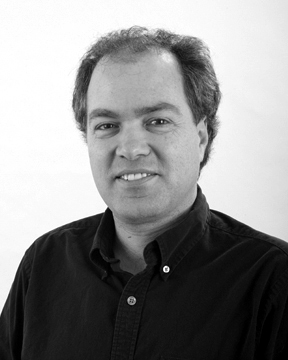}}]{Scott Jordan}
(Member, IEEE) received the B.S./A.B., the M.S., and Ph.D. degrees from the University of California, Berkeley, in 1985, 1987, and 1990, respectively. From 1990 until 1999, he served as a faculty member at Northwestern University. Since 1999, he has served as a faculty member at the University of California, Irvine. In 2006, he served as an IEEE Congressional Fellow, working in the United States Senate on communications policy issues. In 2014-2016, he served as the Chief Technologist at the Federal Communications Commission, advising on technological issues across the Commission, including the 2015 Open Internet Order and the 2016 Broadband Privacy Order. His current research interests are Internet policy issues, including net neutrality, privacy, interconnection, data caps, zero rating, and device attachment.
\end{IEEEbiography}

\end{document}